\documentclass[twocolumn,showpacs,pra,superscriptaddress,longbibliography]{revtex4-2}
\usepackage{float}
\usepackage{times}
\usepackage{amsmath}
\usepackage{amssymb}
\usepackage{graphicx}
\usepackage{physics}
\usepackage[unicode]{hyperref}
\hypersetup{
   unicode=true,          
   plainpages=false,
   colorlinks=true,       
   citecolor=blue,        
   urlcolor=blue
}
\usepackage[caption=false,position=top,singlelinecheck=off,justification=raggedright]{subfig}
\usepackage{color}
\usepackage{pst-node}

\begin{document}
\title{Impact of quantum noise on phase transitions in an atom-cavity system with limit cycles}

\author{Richelle Jade L. Tuquero}
\affiliation{National Institute of Physics, University of the Philippines, Diliman, Quezon City 1101, Philippines}

\author{Jayson G. Cosme}
\email[Corresponding author: ]{jcosme@nip.upd.edu.ph}
\affiliation{National Institute of Physics, University of the Philippines, Diliman, Quezon City 1101, Philippines}

\date{\today}
\begin{abstract}
Quantum fluctuations are inherent in open quantum systems and they affect not only the statistical properties of the initial state but also the time evolution of the system. Using a generic minimal model, we show that quantum noise smoothens the transition between a stationary and a dynamical phase corresponding to a limit cycle (LC) in the semiclassical mean-field approximation of a generic open quantum system. Employing truncated Wigner approximation, we show that the inherent quantum noise pushes the system to exhibit signatures of LCs for interaction strengths lower than the critical value predicted by the standard mean-field theory, suggesting a noise-induced emergence of temporal ordering. Our work demonstrates that the apparent crossover-like behavior between stationary phases brought by finite-size effects from quantum fluctuations also apply to transitions involving dynamical phases. To demonstrate this on a specific physical system, we consider a transversely pumped atom-cavity setup, wherein LCs have been observed and identified as continuous time crystals. We compare the oscillation frequencies of the LCs in the one-dimensional (1D) and two-dimensional regimes, and find that the frequencies have larger shot-to-shot fluctuations in 1D. This has an important consequence in the effectiveness of entrainment of LCs for a periodically driven pump intensity or light-matter coupling strength.
\end{abstract}

\maketitle

\section{Introduction}
Typically, noise is expected to inhibit well-defined oscillatory response in complex systems. Nevertheless, it has been suggested that noise could induce order, instead of the expected chaotic behavior \cite{Matsumoto1983}.
For example, phase synchronization in systems with common noise \cite{Zhou2002} and uncorrelated noise \cite{Nicolaou2020} has been observed.
Moreover, applied noisy drive has been shown to produce nontrivial ordered phases in interacting quantum systems \cite{Moon2024, Zhao2023} and, in other cases, can have the opposite effect of destabilizing an existing limit cycle (LC) or continuous time crystals (CTCs) \cite{Kongkhambut2022}. These LCs are characterized by oscillating observables at a well-defined frequency even though the underlying equations describing the system's time evolution is manifestly time-independent and they can be identified as CTCs in many-body systems. In Ref.~\cite{Zilberberg2023}, fluctuations corresponding to the intrinsic quantum property and coupling to an external environment, and their role on the stability of discrete time crystals have been explored. In this work, we focus on LCs and CTCs as we aim to understand the role of such quantum noises in the emergence of LCs and their dynamical control \cite{Kongkhambut2022, Piazza2015, Kessler2019, Kessler2020, kongkhambut2024}. 

The emergence of LCs in various many-body quantum systems is widely studied, especially in the context of time crystals \cite{Kongkhambut2022, Krishna2023, Greilich2024, Carraro-Haddad2024}. In particular, for an atom-cavity system \cite{Kongkhambut2022}, artificial noise applied to the light-matter interaction strength is found to destabilize the LC, which then begs the question if the inherent fluctuations associated with the photon dissipation in the cavity \cite{Ritsch2013, Mivehvar2021} can similarly inhibit the emergence of LCs in such a system. Furthermore, investigating the precise role of quantum noise on phase transitions between static and dynamical phases is of fundamental importance as it is known that quantum noise in finite systems modify the transition between stationary states, defined by time-independent order parameters, to appear more crossover-like \cite{Polkovnikov2004,Dagvadorj2015,Damanet2019,Orioli2022}. Nevertheless, exploring the precise role of quantum noise can be experimentally challenging as the inherent noise cannot be easily controlled in experiments. In contrast, a theoretical approach such as the truncated Wigner approximation (TWA) \cite{Polkovnikov2010} allows for the possibility of choosing which noise to be included in the theoretical description of a given system. Beyond the mean-field approximation of the quantum dynamics using the TWA has been shown to successfully capture the correct qualitative behavior of time crystals observed in atom-cavity experiments \cite{Cosme2018, Tuquero2022, Cosme2019}.

In this paper, we investigate the effects of quantum noise on phase transitions in a minimal model for an open quantum system that hosts transitions involving stationary and dynamical phases. To this end, we employ both mean-field simulations and the TWA, in which we sample the quantum noise of the initial state and the stochastic noise in time due to the dissipation. 
Our main result suggests a smoothening of the phase transition, implying that signatures of dynamical phases, which are otherwise accessible only for sufficiently strong interactions according to a mean-field (MF) treatment of a system, may emerge in individual trajectories or single-shot time traces for weak interactions. We discuss the implications of our results in the context of light-matter setups operating in the mesoscopic regime of few particles, wherein the fluctuations either from quantum or thermal fluctuations become important as experimentally realized in Ref.~\cite{Ho2024}.
Moreover, we demonstrate the general findings of our minimal model in the specific setup of a transversely pumped atom-cavity system and we investigate the consequence of quantum fluctuations on the entrainment of LCs \cite{kongkhambut2024} in the one-dimensional (1D) and two-dimensional (2D) regimes of this platform.

This paper is structured as follows. We first discuss a minimal model capturing the essential dynamics in the atom-cavity system and related dissipative bosonic systems in Sec.~\ref{sec:minsystem}. Then, we investigate the transversely pumped atom-cavity system in Sec.~\ref{sec:system}. We outline the specific atom-cavity model and corresponding simulation methods beyond the standard mean-field approximation in Sec.~\ref{sec:modelsims}. We present the mean-field phase diagrams for the 1D model in Sec.~\ref{sec:1D2D}. In Sec.~\ref{sec:noise}, we employ the TWA to analyze the qualitative and quantitative effects of quantum noise. We compare the entrainment of LCs for the 1D and 2D atom-cavity systems in Sec.~\ref{sec:entrainment}. Lastly, we conclude the paper in Sec.~\ref{sec:conc}.

\section{Minimal dissipative System and key results}\label{sec:minsystem}

\subsection{Minimal model}

{To highlight the key results of this work, we find it insightful to consider a minimal model that captures the transitions between different phases in the atom-cavity system and related dissipative coupled bosonic systems. Specifically, we study in such a minimal model the effects of quantum fluctuations on the transitions involving static and dynamical phases.}

\begin{figure}[!htb]
	\centering
	\includegraphics[width=1\columnwidth]{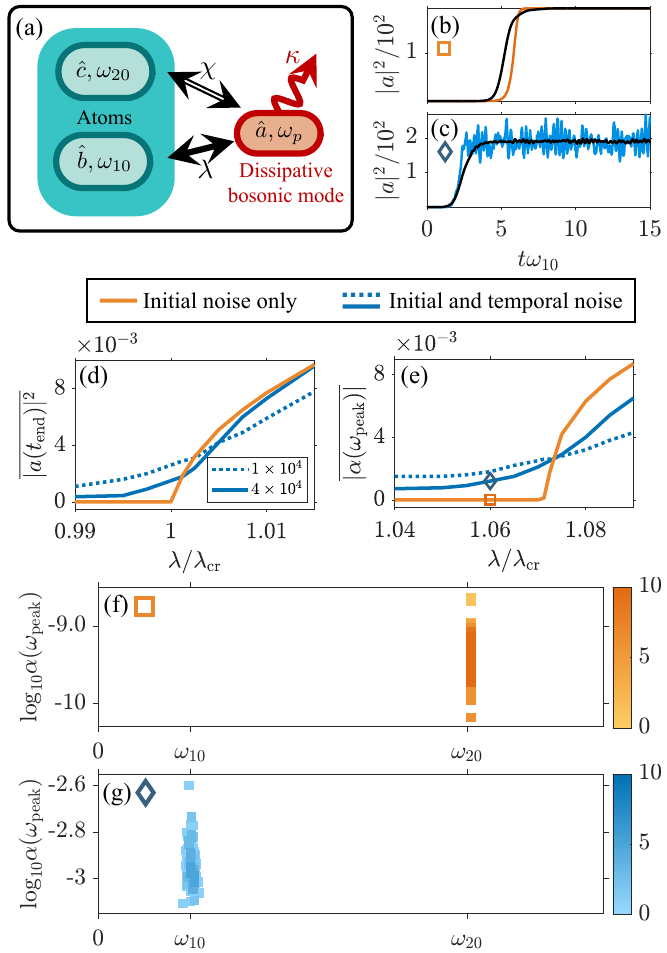}
	\caption{(a) Sketch of the minimal model. Exemplary dynamics of $|a|^2$ for simulations with (b) only the initial noise and (c) both initial and temporal noises. The black curve corresponds to the ensemble average while the light curves correspond to results of a single trajectory. (d) Ensemble averaged $|a|^2$ taken at the final simulation time for different $\lambda$. The legend indicates the number of particles $N$. (e) Ensemble averaged value of the Fourier transform of $|a|^2$ at the peak frequency of each trajectory as a function of $\lambda$. The square and diamond mark the $\lambda$ used for the dynamics in panels (b) and (c). Oscillation amplitude of each trajectory as a function of the corresponding peak frequency with (f) only the initial noise and (g) both initial and temporal noises. The color intensity denotes the number of counts inferred from a histogram of the oscillation amplitude and peak frequency. The $\lambda$ for panels (f) and (g) is marked in (e).}
	\label{fig:minmodel} 
\end{figure} 

{
In Ref.~\cite{Skulte2024}, it was demonstrated that the atom-cavity system with transverse pumping can be mapped onto a simple few-mode dissipative model with a Hamiltonian given by
\begin{align}
\label{eq:minH}
\hat{H} &= \omega_p \hat{a}^\dagger \hat{a} + \omega_{10} \hat{b}^\dagger \hat{b}+\omega_{20} \hat{c}^\dagger \hat{c} +\lambda \left(\hat{a}^\dagger+\hat{a} \right)\left(\hat{b}^\dagger+\hat{b} \right) \\ \nonumber
&\qquad + \chi\sqrt{N} \hat{a}^\dagger \hat{a} \left(\hat{c}^\dagger+\hat{c} \right).
\end{align}
A sketch of the minimal model is shown in Fig.~\ref{fig:minmodel}(a).
Here, $\omega_p$ is the characteristic frequency of the dissipative bosonic mode, i.e., the cavity photons in an atom-cavity platform, and the corresponding bosonic creation and annihilation operators are $\hat{a}^\dagger$ and  $\hat{a}$, respectively. The dissipation in this bosonic mode is characterised by the decay rate $\kappa$. The amplitude of this mode is coupled to the amplitude of a bosonic mode with frequency $\omega_{10}$ represented by the $\hat{b}$-operators. The strength of this interaction is $\lambda$. The last line in Eq.~\eqref{eq:minH} represents the density-dependent interaction with strength $\chi$ between the dissipative mode and a bosonic mode with frequency $\omega_{20}$ corresponding to the $\hat{c}$-operators. We note that the interaction terms in Eq.~\eqref{eq:minH} do not conserve the number of bosons in any of the modes. Nevertheless, we assume that there are $N$ particles initially in the ground state in the nondissipative sector and, in this scenario, the modes associated with the $\hat{b}$- and $\hat{c}$-operators represent excitations out of this state.
}

{
We note that this model applies to a broad class of systems that involves at least three coupled bosonic modes with dissipation and certain types of interactions. An example of the amplitude dependent interactions is the dipole-dipole couplings in light-matter systems. The Hamiltonian for the standard Dicke model under the Holstein-Primakoff transformation, for example, is recovered from Eq.~\eqref{eq:minH} by neglecting the bosonic mode associated with the $\hat{c}$-operators or setting $\omega_{20}=\chi=0$. 
The term proportional to $\chi$ in Eq.~\eqref{eq:minH}, which represents a coupling between the amplitude of a bosonic mode and the density of the dissipative boson, is the key ingredient in the emergence of a LC for sufficiently strong $\lambda$ \cite{Skulte2024}. Applying a mean-field approximation, which effectively transforms the quantum operators to $c$-numbers, i.e., $\langle \hat{a} \rangle \to \tilde{a}$, $\langle \hat{b} \rangle \to \tilde{b}$, and $\langle \hat{c} \rangle \to \tilde{c}$, we obtain the following set of equations of motion (EOM) from the Heisenberg-Langevin equations
\begin{align}\label{eq:eoms3mode}
\frac{{\partial} \tilde{a}}{{\partial} t} &= -i \left[ \omega_p-i\kappa+\chi\sqrt{N} \left(\tilde{c}+\tilde{c}^* \right) \right] \tilde{a}-i\lambda \left(\tilde{b}+\tilde{b}^* \right) + \tilde{\xi},  \\ \nonumber
\frac{{\partial} \tilde{b}}{{\partial} t} &= -i \omega_{10} \tilde{b} -i \lambda  \left(\tilde{a}+\tilde{a}^* \right), \\ \nonumber
 \frac{{\partial} \tilde{c}}{{\partial} t} &= -i \omega_{20} \tilde{c} -i \chi \sqrt{N} \tilde{a}^* \tilde{a}~,
\end{align}
where $\tilde{\xi}$ is the temporal noise obeying $\langle \tilde{\xi}^*(t)\tilde{\xi}(t')\rangle = \kappa \delta(t-t')$. The temporal noise $\tilde{\xi}$ ensures that the correct commutation relations for the dissipative boson are preserved, and thus, is of purely quantum origin \cite{Gardiner,Gardiner1986,Clerk2010}. However, this can be mimicked by a noise drive of the $a$-mode corresponding to an additional term in the Hamiltonian given by $\hat{H}_\mathrm{noise} = i\tilde{\xi}(\hat{a}^\dagger - \hat{a})$, which in such a case would allow for an independent tunability of the dissipation and fluctuation strengths.

To investigate the behavior for varying $N$ and to identify the correct thermodynamic limit, we introduce the rescaled quantities $a \equiv \tilde{a}/\sqrt{N}$, $b \equiv \tilde{b}/\sqrt{N}$, $c \equiv \tilde{c}/\sqrt{N}$, and ${\xi}\equiv \tilde{\xi}/\sqrt{N}$. This leads to the following rescaled EOM
\begin{align}\label{eq:reseoms3mode}
\frac{{\partial} a}{{\partial} t} &= -i \left[ \omega_p-i\kappa+\chi {N} \left(c+c^* \right) \right] a-i\lambda \left(b+b^* \right) + \xi,  \\ \nonumber
\frac{{\partial} b}{{\partial} t} &= -i \omega_{10} b -i \lambda  \left(a+a^* \right), \\ \nonumber
 \frac{{\partial} c}{{\partial} t} &= -i \omega_{20} c -i \chi {N} a^* a~.
\end{align}
The above equations are unchanged for varying $N$ if $\chi N$ is fixed to a constant, which means that the mean-field results in the thermodynamic limit of $N \to \infty$ is obtained by solving Eq.~\eqref{eq:reseoms3mode} with $\chi N = \mathrm{const.}$ and neglecting the stochastic noise, i.e., $\xi = 0$.
}

{To explore the effects of quantum fluctuations, we use the TWA. The TWA includes the leading order quantum corrections to the MF theory \cite{Polkovnikov2010, Carusotto2013, Blakie2008}. Within the framework of TWA, the quantum dynamics is approximated using an ensemble of stochastic trajectories, which sample the initial quantum noise and the stochastic noise associated with the dissipation, in the case of an open system. In theory, we can freely choose which source of quantum noise to include in our TWA simulations. We assume that the system starts in the ground state such that the bosonic modes in our model are initialized as a vacuum state, the corresponding Wigner distribution of which is a Gaussian function \cite{Polkovnikov2010}. We sample this quantum noise by solving Eq.~\eqref{eq:reseoms3mode} using an ensemble of randomized initial states sampled using $\tilde{f} = \left(\eta_{\tilde{f},1} + \eta_{\tilde{f},2} \right)/2$, where $\tilde{f} \in \{\tilde{a},\tilde{b},\tilde{c}\}$ and $\eta_{\tilde{f},i}$ are real random numbers taken from a normal distribution, i.e., $\langle \eta_{\tilde{f},i} \rangle=0$ and $\langle \eta_{\tilde{f},i} \eta_{\tilde{f'},j} \rangle=\delta_{\tilde{f}\tilde{f'}}\delta_{ij}$. For the rest of this work, we refer to this type of quantum noise as the \textit{initial noise} in the system. Distinct from this type of noise is the stochastic noise in time, $\tilde{\xi}$, which we call as the \textit{temporal noise}. 
}

\subsection{Key findings}

{In the following, we numerically solve Eq.~\eqref{eq:reseoms3mode} using 100 trajectories, which sample either (i) just the initial noise or (ii) both the initial and temporal noises, for the experimentally relevant choice of parameters $\kappa=\omega_p=\omega_{10}$, $\omega_{20} = 4\omega_{10}$, and $N\chi = 11$ \cite{Skulte2024}. To solve the stochastic differential equations, we utilize the predictor-corrector method, otherwise known as Heun's method, with a time step of $10^{-3}/\omega_{10}$.
The minimal model possesses a critical point at $\lambda_\mathrm{cr} = [(\kappa^2 + \omega^2_p)\omega_{10}/4\omega_p]^{1/2}$ separating the normal phase (NP) defined by zero steady-state occupation in all of the modes and a stationary phase characterized by nonzero occupation of the dissipative mode $|a|^2>0$, which we dub as the superradiant (SR) phase due to its connection with the phase of the same name in the Dicke model and the atom-cavity system \cite{DickeModel}. For even larger $\lambda$, the SR phase becomes unstable towards the formation of a LC characterised by oscillations at a single dominant frequency \cite{Skulte2024}. The behavior of the dynamics of $|a|^2$ can be used to distinguish between the different phases in the system. To illustrate this, we present in Fig.~\ref{fig:minmodel}(b) an example of the dynamics of $|a|^2$ in the SR phase, which shows that  $|a|^2$ approaches a constant value for long times. 
In contrast, the LC phase exemplified in Fig.~\ref{fig:minmodel}(c) exhibits oscillations at a level of a single trajectory in our TWA simulations, which due to the presence of temporal noise will have additional subdominant frequencies leading to a slightly irregular behavior compared with their purely mean-field counterparts \cite{Skulte2024}. Note that the ensemble averaged $|a|^2$ appears constant for long times in the black curve in Fig.~\ref{fig:minmodel}(c) as a consequence of the spontaneous symmetry breaking of time-translation symmetry in a CTC, leading to the onset of oscillations being chosen randomly for each trajectory \cite{Kongkhambut2022}.
}

{For the transition between the NP and SR phase, the long-time ensemble averaged photon occupation approximated here by $\overline{|a(t_\mathrm{end})|^2}$, where $t_\mathrm{end}$ is the final simulation time and the overline denotes the average over 100 trajectories, is a useful order parameter. If only the initial noise is included as done in the light curve in Fig.~\ref{fig:minmodel}(d), the order parameter $\overline{|a(t_\mathrm{end})|^2}$ follows the expected mean-field behavior that there is a kink at the critical point separating the NP and SR phase, which is reminiscent of a pitchfork bifurcation \cite{Strogatz_nonlinear_2019,DickeModel}. Quantum fluctuations due to finite-size effects captured by the temporal noise within the TWA lead to the smoothening of the transition making it crossover-like as shown in the dark curves in Fig.~\ref{fig:minmodel}(d) for particle numbers $N=10^4$ and $N=4\times 10^4$. This is consistent with the smoothening of the transitions predicted in the atom-only Dicke model \cite{Damanet2019} and is expected to be a generic feature of quantum effects on transitions between stationary phases such as those in Refs.~\cite{Polkovnikov2004,Dagvadorj2015,Orioli2022}. 
}

{
A unique feature of our model in Eq.~\eqref{eq:minH} is the presence of a transition between stationary and dynamical phases, i.e., the SR-LC transition. Thus, it is natural to ask how quantum noise affects such a transition. To this end, we obtain the Fourier transform of $|a|^2$ over a duration of $\tau = 20/\omega_{10}$ and take the ensemble-average value at the most dominant frequency $\overline{\alpha(\omega_\mathrm{peak})}$. $\overline{\alpha(\omega_\mathrm{peak})}$ corresponds to the oscillation amplitude at the frequency $\omega_\mathrm{peak}$. We present a comparison between the scenarios with only the initial noise and both initial and temporal noises in Fig.~\ref{fig:minmodel}(e). The mean-field prediction for the SR-LC transition for the set of parameters used in Fig.~\ref{fig:minmodel}(e) is $\lambda = 1.07 \lambda_\mathrm{cr}$. Thus, the expected steady state for $\lambda=1.06\lambda_\mathrm{cr}$ is an SR phase, which is indeed confirmed by the small oscillation amplitude recorded for the case with only the initial noise marked as the square in Fig.~\ref{fig:minmodel}(e). As a matter of fact, Fig.~\ref{fig:minmodel}(b) is the corresponding dynamics of $|a|^2$ for $\lambda=1.06\lambda_\mathrm{cr}$ when only the initial quantum noise is present. Nevertheless, despite the system being in an SR phase, it is interesting to point out that our Fourier analyses still detected oscillations close to the transition frequency between the ground and second-excited states of the atoms, $\omega_{20}$, albeit with almost negligible oscillation amplitudes $\overline{\alpha(\omega_\mathrm{peak})} \approx 10^{-10}-10^{-9}$ as depicted by the histogram in Fig.~\ref{fig:minmodel}(f). Here, the intensity of the color bar represents the number of trajectories following the binning procedure for the histogram of the oscillations amplitude and peak frequency. We see later that this behavior manifests in the atom-cavity system as transient oscillations at high frequencies close to the available momentum excitations.
}

{Interestingly, the inclusion of temporal noise leads to a smoothening of the SR-LC phase transition, similar to the one observed in the NP-SR transition, as demonstrated by the dark curves in Fig.~\ref{fig:minmodel}(e). The crossover-like behavior becomes more prominent as the quantum fluctuations increase with decreasing particle number. An interesting consequence of this smoothening of the transition is an earlier onset of the LC phase at a lower $\lambda$ compared with the mean-field prediction. This can be seen in the histogram in Fig.~\ref{fig:minmodel}(g), wherein all TWA trajectories exhibit oscillations at a higher amplitude and, more importantly, at the expected LC frequency of $\omega_{10}$ \cite{Skulte2024} compared with their counterparts when only the initial noise is present. Comparing Figs.~\ref{fig:minmodel}(f) and \ref{fig:minmodel}(g), we emphasise that the oscillation amplitudes are different by several orders of magnitude for the two cases. Our results suggest an interpretation that the temporal noise from a purely a quantum origin, namely the preservation of bosonic commutation relations, promotes the onset of the LC phase, which is a phenomenon akin to a noise-induced Hopf bifurcation \cite{Gao2002}. 
This is corroborated by the exemplary dynamics in Fig.~\ref{fig:minmodel}(c) depicting an oscillating $|a|^2$ in a single TWA trajectory for a choice of $\lambda=1.06\lambda_\mathrm{cr}$ that is supposedly on the SR side of the SR-LC phase transition according to mean-field theory.
The smoothening of the SR-LC transition leading to an apparent noise-induced LC in the quantum regime is the main finding of this work.} 

\begin{figure}[!htpb]
	\centering
	\includegraphics[width=1\columnwidth]{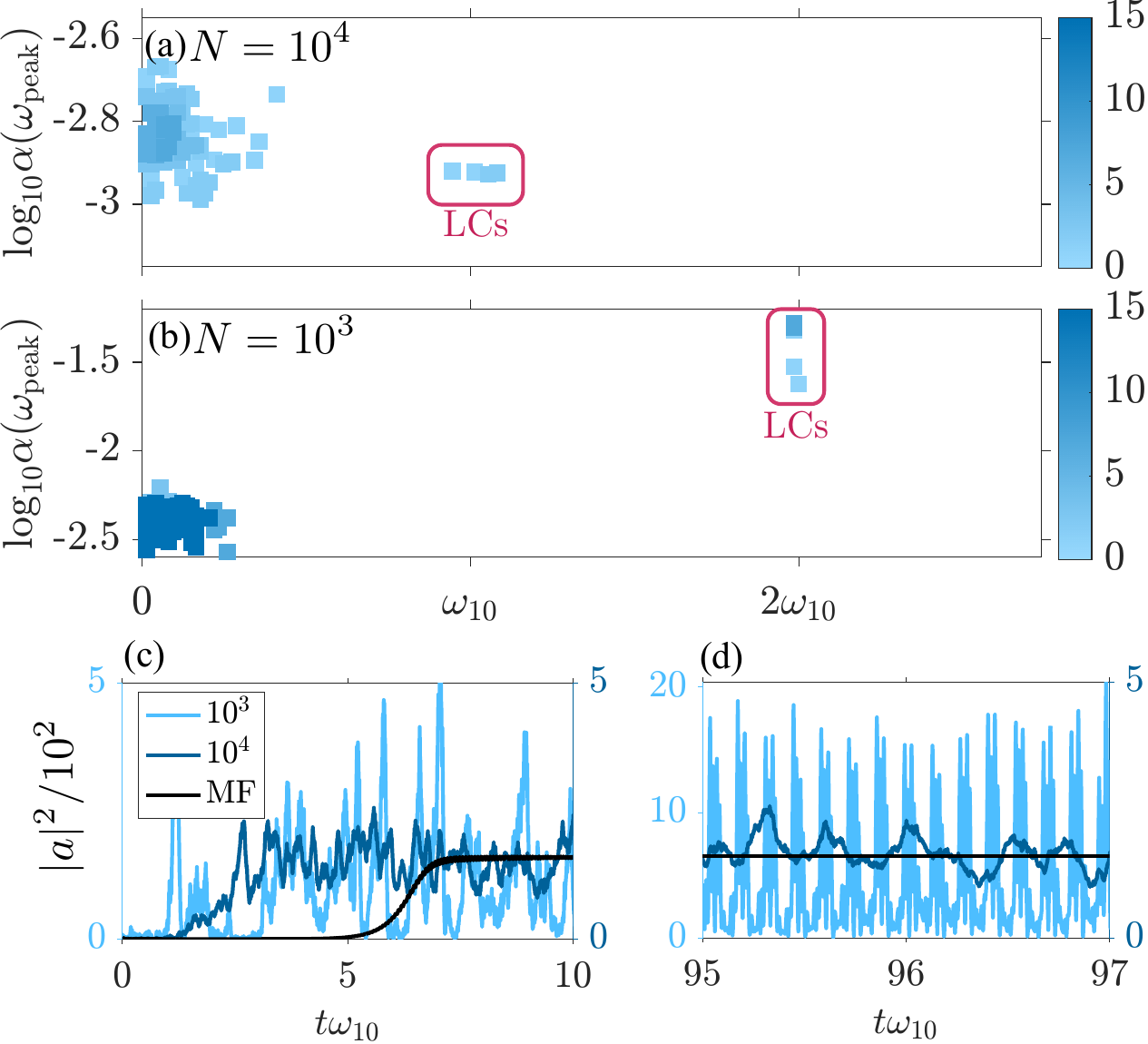}
	\caption{(a), (b) Similar to Fig.~\ref{fig:minmodel}(g) but for $\lambda = 1.04\lambda_\mathrm{cr}$ and $N$ as indicated in the legends. LC trajectories are enclosed in rectangles. (c), (d) Single-shot simulations representing LC trajectories in (a) and (b), i.e., solutions with oscillation frequency close to $\omega_{10}$ in (a) and $2\omega_{10}$ in (b).}
	\label{fig:N} 
\end{figure}

We now compare the general predictions of our minimal model with the experimental observations related to phase transitions in a related light-matter system operating at the mesoscopic regime of few atoms \cite{Ho2024}. We obtain the histogram of the oscillation amplitude and frequency for a particle number small enough for some of the TWA trajectories to start showing signs of the ``switching" dynamical behavior predicted in Refs.~\cite{Schutz2015,Stitely2020} and observed in Ref.~\cite{Ho2024}, which we identify in our system to start occurring for $N=10^3$. The switching dynamics is characterized by a change in the sign of the order parameter, which here is given by $\mathrm{Re}(a)$, at random points in time \cite{Schutz2015,Stitely2020}. In Figs.~\ref{fig:N}(a) and \ref{fig:N}(b), we present the histograms close to the NP-SR critical point, $\lambda = 1.04\lambda_\mathrm{cr}$, for $N=10^4$ and $N=10^3$, respectively.  In both cases, a large fraction of single-shot simulations remains in the SR phase with oscillation frequencies close to $\omega=0$ and only a small fraction $\approx 10\%$ leads to a LC.  However, there is a qualitative difference in the type of LC solutions between the mesoscopic ($N=10^3$) and macroscopic ($N=10^4$) regimes: the LCs for $N=10^3$ oscillate at $2\omega_{10}$, instead of the MF prediction of $\omega_\mathrm{LC} = \omega_{10}$ that the LCs for macroscopic particle number $N\geq10^4$ follow. 
The qualitative change in behavior is evident in the individual limit cycle (LC) trajectories shown in Figs.~\ref{fig:N}(c) and \ref{fig:N}(d). The LCs for $N=10^3$ exhibit pulsating dynamics at nearly twice the frequency of the oscillations observed for $N=10^4$.

\begin{figure}[!htpb]
	\centering
	\includegraphics[width=1\columnwidth]{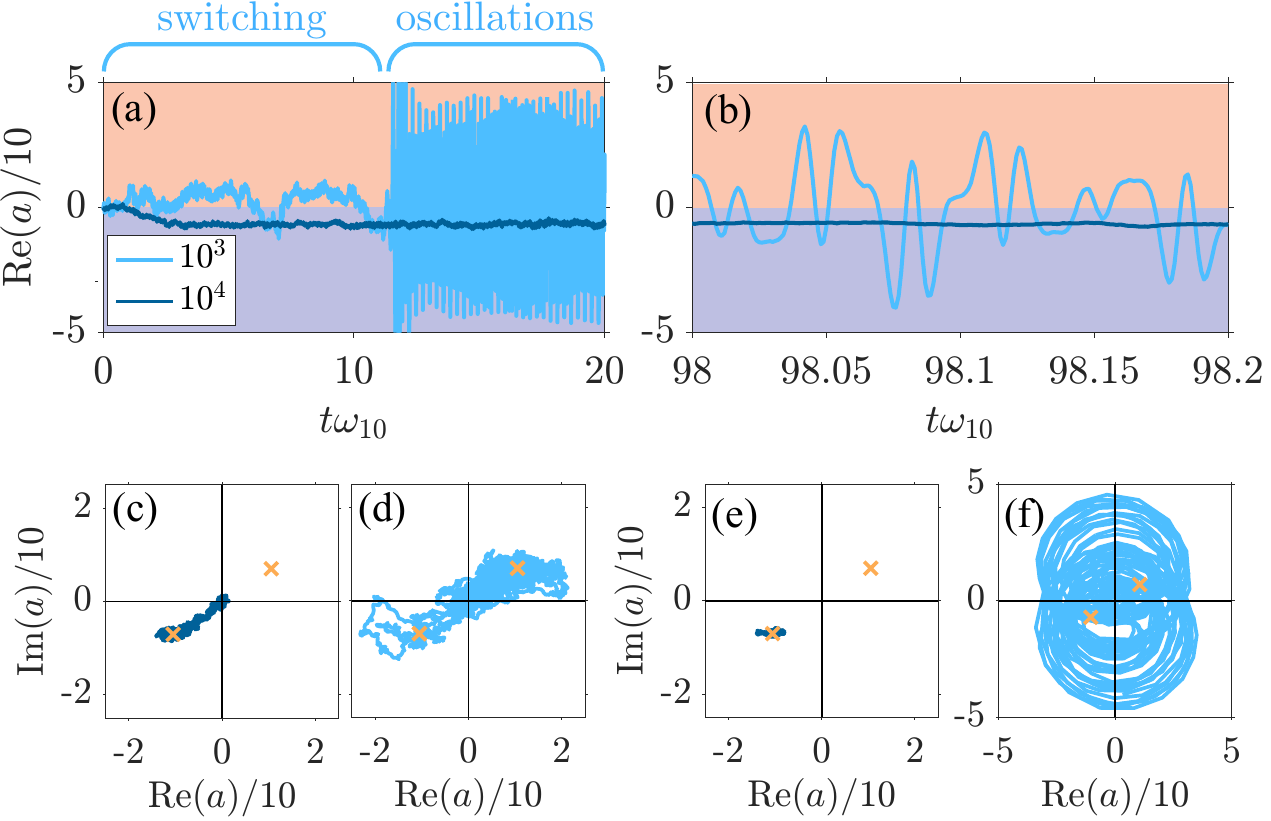}
		\caption{Dynamics of the order parameter $\mathrm{Re}(a)$ for individual LC trajectories in the (a) short-time and (b) long-time limits. The shaded areas correspond to the two possible symmetry broken dynamics, $\mathrm{Re}(a)<0$ and $\mathrm{Re}(a)>0$. Trajectories in phase space spanned by $\mathrm{Re}(a)$ and $\mathrm{Im}(a)$ for (c), (d) $t\omega_{10}<10$ and (e), (f) $t\omega_{10}>98$. The crosses mark the symmetry broken SR fixed points according to MF. The parameters are the same as in Fig.~\ref{fig:N}.}
	\label{fig:phaseportrait} 
\end{figure} 

The dynamics of the order parameter $\mathrm{Re}(a)$ is insightful in distinguishing the unique behaviors in the mesoscopic and macroscopic regimes. As mentioned earlier for $N=10^4$ and now explicitly shown in Fig.~\ref{fig:phaseportrait}(a), we find that all trajectories, both SR and LC solutions, demonstrate symmetry-breaking dynamics, in which the sign of $\mathrm{Re}(a)$ does not change over time. On the other hand, in the mesoscopic regime, a single trajectory may exhibit not only switching dynamics, which is the expected behavior in the quantum regime of a SR phase, but also sign-changing oscillations of $\mathrm{Re}(a)$ as exemplified in the time trace for $N=10^3$ in Figs.~\ref{fig:phaseportrait}(a) and \ref{fig:phaseportrait}(b). Similar sign-changing oscillations have been experimentally observed in a minority of the single-shot time traces in Ref.~\cite{Ho2024}, wherein their existence was attributed to a difference in energies of a dominant mode and the gap separating the symmetry-broken SR states. Here, our semiclassical perspective on the trajectories used to approximate the quantum dynamics in the system allows for an interpretation of the switching and oscillating responses using the language of nonlinear dynamics. As $N$ decreases, the temporal noise effectively increases leading to strong fluctuations in time of the individual trajectories. For strong enough fluctuations, a solution can be pushed to the other symmetry broken state, thereby realising the switching dynamics. This can be seen from the comparison of phase space dynamics for $N=10^4$ and $N=10^3$ in Figs.~\ref{fig:phaseportrait}(c) and \ref{fig:phaseportrait}(d), respectively, in which strong temporal fluctuations lead to switching dynamics in Fig.~\ref{fig:phaseportrait}(d). An even stronger temporal perturbation can occur randomly in time and it can drive the system into a completely different type of attractor, e.g., the LC phase shown in Fig.~\ref{fig:phaseportrait}(f). This behavior is distinct from the type of LC in the macroscopic regime [cf. Fig.~\ref{fig:phaseportrait}(e)] corresponding to small amplitude oscillations around a fixed point or symmetry broken state at an oscillation frequency close to that of a stable LC in the thermodynamic limit. Thus, we provide an alternative semiclassical interpretation for the observation of oscillating dynamics in the mesoscopic regime, which is the emergence of noise-induced LCs in a system that potentially hosts a dynamical phase. We note that a beyond MF analysis on the stability of a spatiotemporal order in a specific configuration of the atom-cavity platform \cite{Nie2024} also predicts the emergence of temporal ordering in a parameter regime identified to be unstable by MF theory \cite{Zhang2022}.

\section{Atom-Cavity System}\label{sec:system}

\subsection{Model and simulation methods}\label{sec:modelsims}

\begin{figure}[!htb]
	\centering
	\includegraphics[width=0.95\columnwidth]{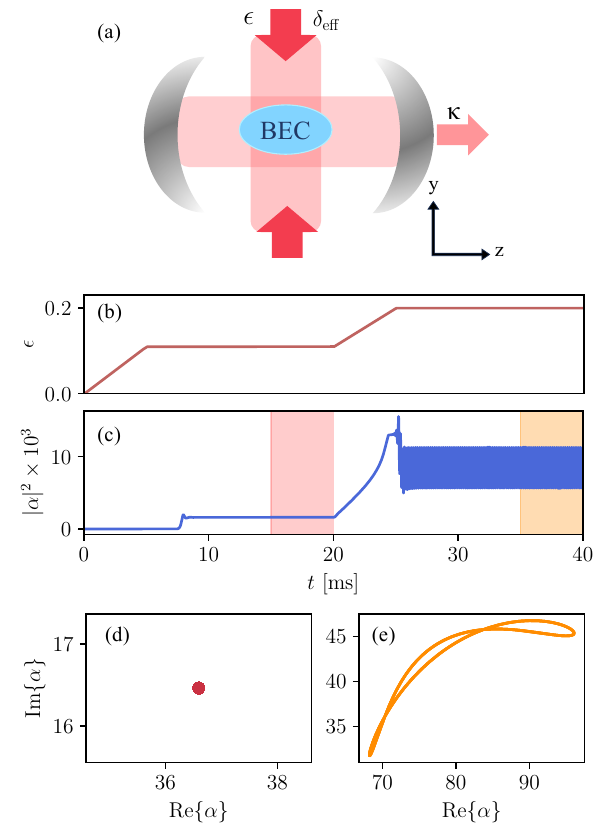}
	\vspace{-0.4cm}
	\caption{(a) Schematic diagram of an atom-cavity system with dissipation rate $\kappa$ and wavelength $\lambda$. The laser pump intensity is $\epsilon$ with effective detuning $\delta_\mathrm{eff}$. To exemplify the dynamics of the superradiant and limit cycle phase, we apply a (b) two-ramp pump protocol. (c) Corresponding dynamics of the intracavity photon number. Long-time trajectories in a complex plane spanned by $\mathrm{Re}\{\alpha\}$ and $\mathrm{Im}\{\alpha\}$ for the (d) Superradiant phase and (e) Limit cycle phase.}
	\label{fig:LCalpha} 
    \vspace{-0.5cm}
\end{figure}

We now explore the manifestation of the findings from the previous section and their consequences in a specific physical setup, which is an atom-cavity system of $^{87}$Rb-atoms forming a Bose-Einstein condensate (BEC) with particle number $N_a = 6.5\times 10^4$ trapped inside a Fabry-P\'{e}rot resonator as shown in Fig.~\ref{fig:LCalpha}(a). A pump beam with intensity $\varepsilon = \epsilon E_\mathrm{rec}$, where $E_\mathrm{rec} = \hbar\omega_\mathrm{rec}$ is the recoil energy, is applied in the direction perpendicular to the cavity axis. The pump has a wavelength $\lambda_p = 792.55$ nm, which means that the wave vector is $k=2\pi/\lambda_p$, and the recoil frequency is $\omega_\mathrm{rec} = 2\pi^2\hbar/m\lambda_p$, where $m$ is the mass of a $^{87}$Rb atom.
Photons leaking out of the cavity is characterised by the cavity photon decay rate $\kappa= 2\pi\times 3.6$ kHz. We use realistic parameters based on experimental realizations in the good cavity regime $\kappa\sim\omega_\mathrm{rec}$ \cite{Kessler2019, Kessler2020, Kessler2021, Kongkhambut2022, kongkhambut2024}. 

The total Hamiltonian for the atom-cavity system consists of the cavity, atom, and atom-cavity interaction terms,
\begin{equation}\label{H}
	\hat{H} = \hat{H}_\mathrm{C}+\hat{H}_\mathrm{A}+\hat{H}_\mathrm{AC}.
\end{equation}
We neglect the atomic interactions, which break the mean-field solvability of this model \cite{Tuquero2022}.
The Hamiltonian for the cavity described by a mode function $\cos(kz)$ is given by
\begin{equation}\label{Hc}
	\hat{H}_\mathrm{C} = -\hbar\delta_\mathrm{C}\hat{\alpha}^\dagger \hat{\alpha}, 
\end{equation}
where $\delta_\mathrm{C}$ is the pump-cavity detuning and $\hat{\alpha}^\dagger$ ($\hat{\alpha}$) is the creation (annihilation) operator for the cavity photon. Note that bare pump-cavity detuning $\delta_\mathrm{C}$ is related to the atom-shifted frequency $\delta_\mathrm{eff}$ via $\delta_\mathrm{C} = \delta_\mathrm{eff} +N_a U_0/2$, where $U_0$ is the lattice depth produced by a single photon in the cavity. The Hamiltonian for the atoms is
\begin{equation}\label{Ha}
	\hat{H}_\mathrm{A} = \int \hat{\Psi}^\dagger(y,z) \left[-\frac{\hbar^2}{2m}\nabla^2+\varepsilon \cos^2(ky)\right]\hat{\Psi}(y,z) \;dy\:dz
\end{equation}
where $\hat{\Psi}(y, z)$ is the bosonic field operator associated with the BEC.
Lastly, we consider the atom-cavity interaction
\begin{align}\label{eq:Hac}
	\hat{H}_\mathrm{AC} &=\int \hat{\Psi}^\dagger(y,z) \;\hbar U_0  \biggl[\cos^2(kz)\hat{\alpha}^\dagger \hat{\alpha} \\ \nonumber
	&+\sqrt{\frac{{\varepsilon}}{\hbar|U_0|}} \cos(kz)\cos(ky)(\hat{\alpha}^\dagger +\hat{\alpha})\biggr]\hat{\Psi}(y,z)\; dy\;dz.
\end{align}

The dynamics of the system is given by the Heisenberg-Langevin equations
\begin{align}\label{eq:eom}
	\frac{\partial}{\partial t}\hat{\Psi} &= \frac{i}{\hbar}[\hat{H}, \hat{\Psi}] \\ 
	\frac{\partial}{\partial t}\hat{\alpha} &= \frac{i}{\hbar}[\hat{H}, \hat{\alpha}]-\kappa \hat{\alpha}+\xi.
\end{align}
where $\xi$ is the stochastic noise due to the cavity dissipation with $\langle \xi^*(t)\xi(t')\rangle = \kappa \delta(t-t')$ \cite{Ritsch2013, Mivehvar2021}. {For the exact mapping between the atom-cavity system parameters and Hamiltonian, and their counterparts in the minimal model in Sec.~\ref{sec:minsystem}, we refer the reader to Ref.~\cite{Skulte2024}.}

We now provide a concise discussion on the truncated Wigner approximation as applied on the atom-cavity model. For a symmetrically ordered expression of the Wigner-Weyl quantization, we can use the Bopp representation to transform the quantum operators to $c$-numbers \cite{Polkovnikov2010}
\begin{gather}
    \hat{\psi}^\dagger \rightarrow \psi^*-\frac{1}{2}\frac{\partial}{\partial\psi},\:
    \hat{\psi} \rightarrow \psi+\frac{1}{2}\frac{\partial}{\partial\psi^*}\\
    \hat{\alpha}^\dagger \rightarrow \alpha^*-\frac{1}{2}\frac{\partial}{\partial\alpha}, \: \hat{\alpha} \rightarrow \alpha+\frac{1}{2}\frac{\partial}{\partial\alpha^*}.
\end{gather}
Using the Moyal product, the leading order of the commutator's Weyl symbol is approximated as
\begin{equation}
    [\hat{O}_1, \hat{O}_2] \approx O_{1, W}\Lambda_c O_{2, W}
\end{equation}
where
\begin{equation}
    \Lambda_c = \sum_k \frac{\overleftarrow{\partial}}{\partial a_k}\frac{\partial}{\partial a_k^*}-\frac{\overleftarrow{\partial}}{\partial a_k^*}\frac{\partial}{\partial a_k}.
\end{equation}

In the 1D limit, the dynamics along the pump direction is frozen $\hat{\Psi}(y,z) \to \hat{\Psi}(z)$, such that Eq.~\ref{Ha} reduces to just the kinetic energy along the cavity direction. Moreover, the atom-cavity interaction Eq.~\ref{eq:Hac} simplifies to only contain $z$-dependent terms, which can be obtained by taking $\cos(ky) \to 1$.
Applying the above transformations on Eq.~\eqref{eq:eom}, the semiclassical EOM for a one-dimensional atom-cavity are \cite{Tuquero2022}
\begin{equation}\label{eom_psi}
\begin{split}
		i\hbar 	\frac{\partial \psi_n}{\partial t}	&=-\frac{\hbar^2}{2m}\nabla^2\psi_n
		+\hbar U_0 \psi_n\biggl[\cos^2(kz_n)\left(|\alpha|^2-\frac{1}{2}\right)\\
        &+\sqrt{\frac{\varepsilon}{\hbar |U_0|}}\cos(kz_n)(\alpha^*+\alpha)\biggr]
\end{split}
\end{equation}
\begin{equation}\label{eom_a}
\begin{split}
i\frac{\partial \alpha}{\partial t} &= -\delta_c\alpha+U_0\sum_n \biggl(|\psi_n|^2-\frac{1}{2}\biggr)\biggl[\cos^2(kz_n)\alpha\\
&+\sqrt{\frac{\varepsilon}{\hbar|U_0|}}\cos(kz_n)\biggr]-i\kappa\alpha+i\xi.
\end{split}
\end{equation}
where $\nabla^2\psi_n$ is resolved using a five-point stencil for approximating the second derivative.

The two-dimensional model, on the other hand, is more conveniently solved in momentum basis. We then expand the atomic field operators according to \cite{Cosme2018,kongkhambut2024,Cosme2019}
\begin{equation}
    \hat{\Psi}(y,z) =\sum_{n,m}\hat{\phi}_{n,m}e^{inky}e^{imkz}
\end{equation}
where $\hat{\phi}_{n,m}^\dagger$ ($\hat{\phi}_{n,m}$) is the bosonic creation (annihilation) operator. Applying the truncated Wigner formalism, the semiclassical EOM in a compact form read
\begin{gather}
    i\hbar\frac{\partial\phi_{n,m}}{\partial t} = \frac{\partial H}{\partial \phi_{n,m}^*}\\
    i\frac{\partial \alpha}{\partial t}= \frac{1}{\hbar}\frac{\partial H}{\partial \alpha^*}-i\kappa\alpha+i\xi.
\end{gather}
The expanded form of the EOM for the two-dimensional system are given by \cite{Cosme2018}
\begin{align}\label{eq:eom2Da}
	i &\frac{\partial \phi_{n,m}}{\partial t} = 
	\omega_{\mathrm{rec}}\left(n^2 + m^2+\frac{U_0}{2\omega_{\mathrm{rec}}}|\alpha|^2-\frac{\epsilon}{2}\right)\phi_{n,m} \\ \nonumber
	&+\frac{U_0}{4}|\alpha|^2(\phi_{n,m-2}+\phi_{n,m+2}) -\frac{\epsilon \omega_{\mathrm{rec}}}{4}(\phi_{n-2,m}+\phi_{n+2,m})\\ \nonumber
	 &+\frac{\sqrt{\epsilon |U_0|\omega_{\mathrm{rec}}}}{2}\mathrm{Re}({\alpha})\biggl[\phi_{n-1,m-1}+\phi_{n+1,m-1} \\ \nonumber
	 & \qquad \qquad \qquad  \qquad +\phi_{n-1,m+1}+\phi_{n+1,m+1}\biggr]
\end{align} 
\begin{align}\label{eq:eom2Db}
	 i &\frac{\partial \alpha}{\partial t} = \left[-\delta_{\mathrm{eff}} + \frac{1}{2}N_aU_0 \sum_{n,m}\mathrm{Re}[\phi_{n,m}\phi^{*}_{n,m+2}]-i\kappa \right]\alpha \\ \nonumber
	 &+\frac{N_a\sqrt{\epsilon |U_0|\omega_{\mathrm{rec}}}}{4}\sum_{n,m}\biggl[\phi_{n,m}(\phi^{*}_{n+1,m+1}+\phi^{*}_{n+1,m-1}) \\ \nonumber
	 & \qquad  \qquad + \phi^{*}_{n,m}(\phi_{n+1,m+1}+\phi_{n+1,m-1})\biggr] + i\xi.
\end{align}

To numerically solve the relevant stochastic differential equations, we use the same Heun's method employed in the minimal model. We consider a time step of $10^{-8}$ s and record the values every 20 steps. {We numerically solve the 1D EOM using 32 grid points within a unit cell spanned by $z\in[0,\lambda_p]$. On the other hand, for the 2D simulations, we use the momentum states spanning $\{p_y,p_z\}\in[-6,6]\hbar k$, i.e., a $13\times 13$ momentum grid. For the initial state, we assume a coherent state for the BEC and a vacuum state for the cavity mode.
}

\subsection{Mean-field results}\label{sec:1D2D}
We first focus on mean-field simulations to serve as a benchmark for TWA simulations.
To introduce the phases found in this configuration of an atom-cavity system, we implement the two-ramp pump protocol shown in Fig.~\ref{fig:LCalpha}(b). In Fig.~\ref{fig:LCalpha}(c), we show the dynamics of the intracavity photon number, $|\alpha|^2$, used to distinguish between different phases in the system. The NP is identified by a zero intracavity photon number. By increasing the pump intensity, the system will transition to an SR phase for $\varepsilon\geq \varepsilon_\mathrm{SR}$, where $\varepsilon_\mathrm{SR}$ is the critical value for the NP-SR transition.
The SR phase is characterized by a nonzero constant intracavity photon number \cite{Ritsch2013, Baumann2010, Nagy2008, Klinder2015} as highlighted in Fig.~\ref{fig:LCalpha}(c). Hence, its long-time trajectory as shown in Fig.\ref{fig:LCalpha}(d) appears as a fixed point in the phase space spanned by the real and imaginary parts of the cavity mode $\alpha$. Increasing the pump intensity further leads to a transition into the LC phase. This dynamical phase is characterized by an oscillating $|\alpha|^2$ at a well-defined limit cycle frequency $\omega_\mathrm{LC}$ as shown in Fig.~\ref{fig:LCalpha}(c). Because it oscillates at a certain frequency, the long-time trajectory in phase space forms a closed loop as presented in Fig.~\ref{fig:LCalpha}(e).

\begin{figure}[!hbt]
	\centering
        \vspace{0.25cm}
	\includegraphics[width=0.95\columnwidth]{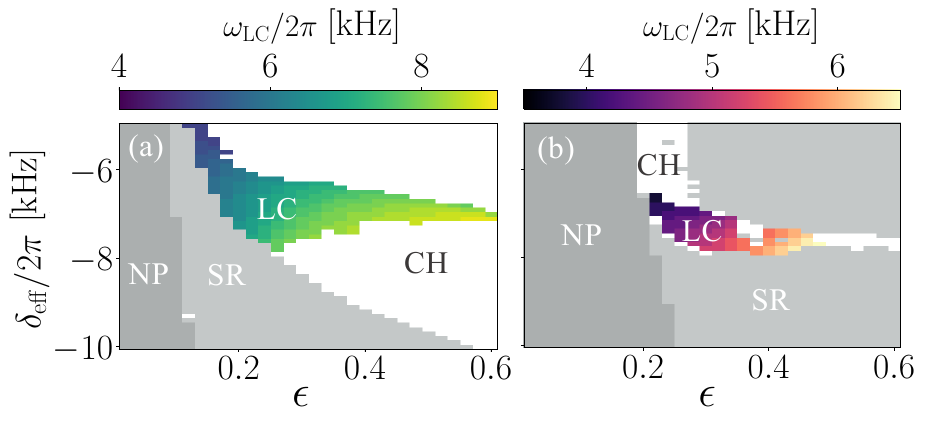}
	\vspace{-0.25cm}
	\caption{Phase diagram of a one-dimensional atom-cavity system when (a) $U_0 = 2\pi\times 0.7$ Hz for the blue-detuned case, and (b) $U_0 = -2\pi\times 0.34$ Hz for the red-detuned case. We show the Normal phase (NP), Superradiant (SR), limit cycle (LC), and Chaos (CH). The LC phase includes the frequencies as given by the color bar on top.}
	\label{fig:1} 
\end{figure}
In Fig.~\ref{fig:1}, we present the phase diagram in the $\delta_\mathrm{eff}$-$\epsilon$ space for a 1D system. We note that the existence of LCs in the 1D regime for blue and red atom-pump detunings was first predicted in Refs.~\cite{Piazza2015} and \cite{Gao2023}, respectively. 
To construct Fig.~\ref{fig:1}, we linearly increase the pump intensity until $\epsilon$ for $t\in [15, 20]~\mathrm{ms}$. Then, we fix $\epsilon$ to a constant value for 15 ms. We consider the time frame of the last 5 ms to obtain the long-time average, maximum value, and the standard deviation of the rescaled peak heights of the cavity occupation. In classifying the dynamics, we apply the following threshold values. 
A cavity occupation dynamics is classified to be that of a LC if the standard deviations is lower than 0.025, and the difference between the maximum and average $|\alpha|^2$ is greater than or equal to 10\% of the average. The SR phase, on the other hand, is identified by having a standard deviation less than 10\% of the average cavity occupation. A response with an average photon occupation less than or equal to $10^{-3}$ is considered as a NP. Lastly, dynamics with standard deviation greater than or equal to 0.025 is determined as a chaotic response (CH). We obtain the LC frequency $\omega_\mathrm{LC}$ from the dominant frequency in the power spectrum dynamics of the intracavity photon number $|\alpha|^2$ for the same time window. Then, $\omega_\mathrm{LC}$ is obtained as the frequency with the highest peak in the power spectrum. 
In Fig.~\ref{fig:1}, both the blue-detuned and red-detuned cases have LCs. 

\begin{figure}[!htb]
	\centering
        \vspace{0.25cm}
	\includegraphics[width=0.98\columnwidth]{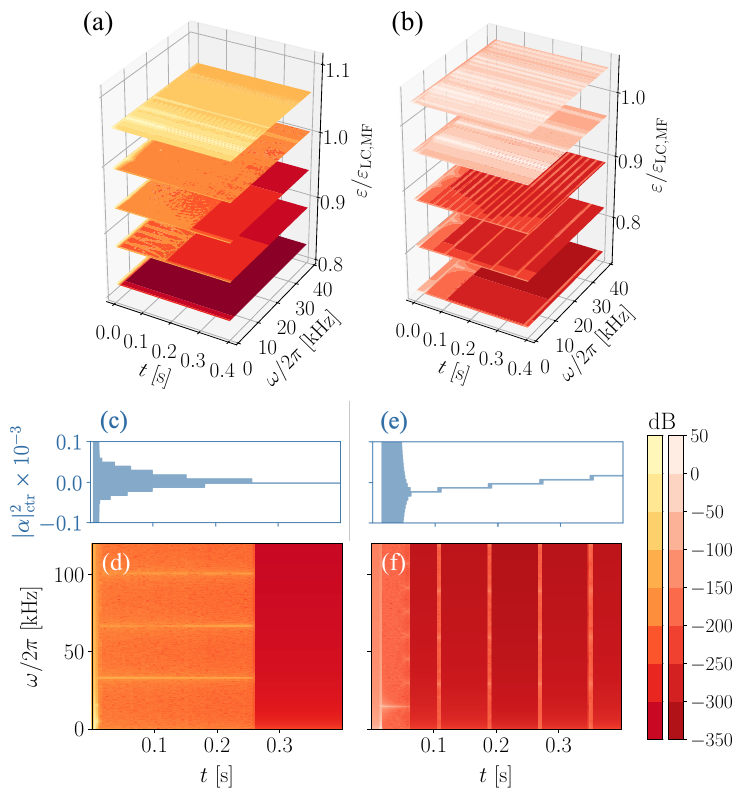}
	\vspace{-0.1cm}
	\caption{Gallery of spectrograms for (a) 1D and (b) 2D atom-cavity systems. Each slice corresponds to a pump intensity rescaled by the pump intensity needed to enter the LC phase according to mean-field theory, $\varepsilon/\varepsilon_\mathrm{LC, MF}$. The effective detuning for 1D and 2D is $\delta_\mathrm{eff} = -2\pi\times 7$ kHz and $\delta_\mathrm{eff} = -2\pi\times 4.7$ kHz, respectively. Intracavity photon dynamics and spectogram for (c-d) 1D with $\varepsilon/\varepsilon_\mathrm{LC, MF}\approx 0.94$, and (e-f) 2D with $\varepsilon/\varepsilon_\mathrm{LC, MF}\approx 0.81$.}
	\label{fig:spec_mf} 
\end{figure} 
In Fig.~\ref{fig:spec_mf}, we further investigate the dynamics of $|\alpha|^2$ by including a gallery of spectrograms for different $\varepsilon/\varepsilon_\mathrm{LC, MF}$, where $\varepsilon_\mathrm{LC, MF}$ is the intensity required to transition from SR to LC according to MF theory. This allows us to take note of how the dynamics change for different intensities in the absence of temporal noise. Accordingly, we expect that the system is in the SR phase when $\varepsilon/\varepsilon_\mathrm{LC, MF}<1.0$, and it is in the LC phase when $\varepsilon/\varepsilon_\mathrm{LC, MF}\geq1.0$. 
For the pump protocol, we linearly increase the pump intensity until the desired $\varepsilon$ for 5 ms, then kept it constant for 395 ms. Afterwards, we obtain the spectrogram of $|\alpha|^2$ using a moving time window of $\approx 3.3$ ms. The intensity in units of dB corresponds to 
$10\log_{10}(P)$ where $P$ is the power spectral density. Hence, the dominant oscillation frequencies of $|\alpha|^2$ are characterized by positive intensity shown as brighter shades in the color bars. Thus, a dominant frequency seen as streak of positive intensity at a constant $\omega$ signifies the appearance of a LC. On the other hand, the SR phase is identified by the lack of nonzero frequency peak in the long-time limit due to the absence of photon oscillations. 

We find that in 1D, as depicted in Fig.~\ref{fig:spec_mf}(a), transient oscillations appear but they eventually vanish for long times as $\varepsilon$ approaches $\varepsilon_\mathrm{LC, MF}$. 
We note for the discussions in the next section that the transient oscillations in Fig.~\ref{fig:spec_mf} persist for $t<100$ ms.
To inspect these dynamics, we show in Fig.~\ref{fig:spec_mf}(c-f), the adjusted intracavity dynamics $|\alpha|^2_\mathrm{ctr} = |\alpha|^2-|\alpha|^2_\mathrm{mean}$ where $|\alpha|^2_\mathrm{mean}$ is the mean value of $|\alpha|^2$ for $t\in[0.1, 0.4]$ s, and its corresponding spectrogram for $\varepsilon/\varepsilon_\mathrm{LC, MF}<1.0$. We find that the amplitude of the $|\alpha|^2_\mathrm{ctr}$ oscillation decreases and eventually vanishes. 
This is also reflected in the features of the spectrogram in Fig.~\ref{fig:spec_mf}(d). It is also striking that a consistent dominant frequency is present, and that it is larger than the expected $\omega_\mathrm{LC}$. 
This can be attributed to excitation of higher momentum modes as inferred from the behavior observed in the minimal model in Fig.~\ref{fig:minmodel}(f).
On the other hand, the 2D system shows a different trend in Fig.~\ref{fig:spec_mf}(b). While we find that the transient oscillations vanish after some time similar to the 1D case, we find that the photon dynamics approach what appears to be plateaus that increase in value as portrayed in Fig.~\ref{fig:spec_mf}(e). Moreover, between these plateaus, we observe short-lived small amplitude oscillations that become more frequent the closer $\varepsilon$ is to $\varepsilon_\mathrm{LC, MF}$. The sharp increase in the photon number appears as streaks at constant times in the spectrogram in Figs.~\ref{fig:spec_mf}(b) and~\ref{fig:spec_mf}(f).
While the appearance of these transient oscillations are interesting, we emphasise that they are difficult to detect both theoretically and experimentally since the corresponding oscillation amplitudes are small as seen in the $y$-axis of Figs.~\ref{fig:spec_mf}(c) and \ref{fig:spec_mf}(f). Furthermore, we see later that they become overshadowed by quantum noise.

To summarize, the mean-field regime provides us with a baseline of the behavior of various phases in the system. We find transient oscillations in the absence of temporal noise for the SR phase. Hence, one may mistakenly identify these transient oscillatory response as LC if the simulation time is not sufficiently long. We will therefore classify such a behavior with an oscillatory response faster than the expected LC frequency as a SR phase. 

\subsection{Truncated Wigner approximation}\label{sec:noise}
Next, we explore the effects of quantum noise from the initial state and temporal noise associated with the cavity dissipation as done in Sec.~\ref{sec:modelsims}.
\begin{figure}[!hbt]
	\centering
	\includegraphics[width=0.97\columnwidth]{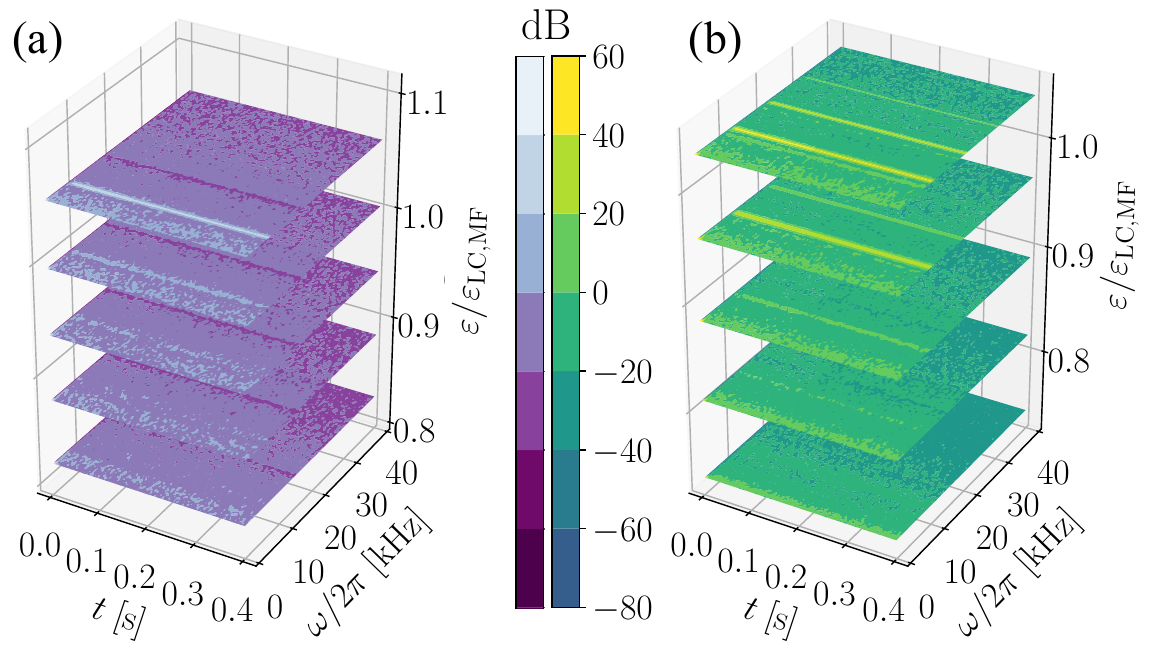}
	\vspace{-0.4cm}
	\caption{Spectrograms for (a) 1D and (b) 2D with the same set of parameters as in Fig.~\ref{fig:spec_mf} but including temporal noise.}
	\label{fig:spec_noise} 
\end{figure}

In Fig.~\ref{fig:spec_noise}, we present the results for the same pump protocol used in Fig.~\ref{fig:spec_mf} albeit now in the presence of temporal noise. Note that we use exactly the same initial state for a fair comparison. As expected when temporal noise is included in the dynamics, we find that the noise broadens the power spectrum, which results to a decrease in the intensities of the dominant frequencies. Interestingly, the transient high-frequency oscillations in the SR phase seen in Fig.~\ref{fig:spec_mf} have vanished. This means that the inherent quantum noise due to dissipation overwhelms the transient oscillations making them unlikely to be observed in actual experiments. Moreover, it also pushes the system to oscillate at a well-defined frequency such that $\omega_\mathrm{peak}\sim\omega_\mathrm{LC, MF}$ as seen from the bright streaks close to the expected LC frequency in Fig.~\ref{fig:spec_noise}. The spectrograms for the SR phases close to the SR-LC transition point already show signs of a LC by having a faint bright streak at the expected LC frequency in Fig.~\ref{fig:spec_noise}(b), for example. 

\begin{figure}[!tbh]
	\centering
	\includegraphics[width=0.97\columnwidth]{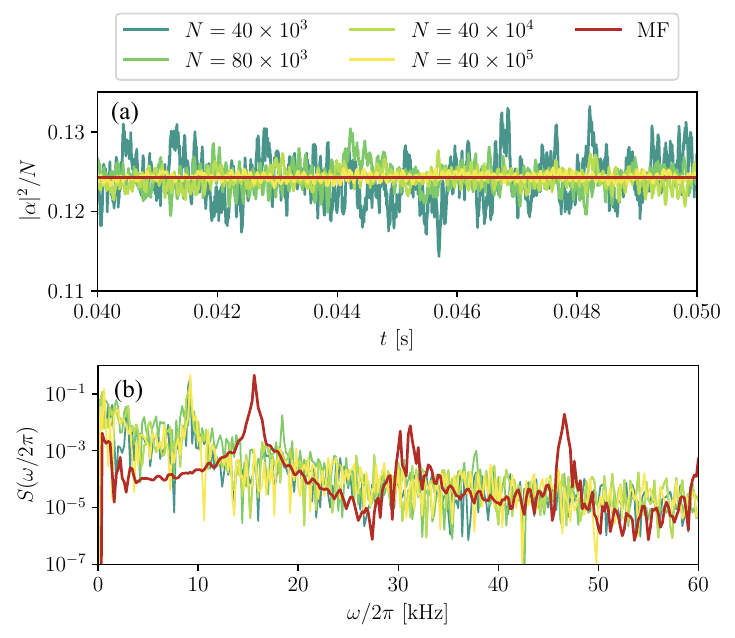}
	\vspace{-0.25cm}
	\caption{(a) The rescaled intracavity photon number by the number of atoms $N$ for MF, and varying $N$ with temporal noise when $\varepsilon/\varepsilon_\mathrm{LC, MF}\approx 0.89$ in 2D. (b) The corresponding normalized power spectrum in logarithmic scale for visibility without the zeroth component.}
	\label{fig:N_cav_fft} 
\end{figure}

In Fig.~\ref{fig:N_cav_fft}, we further investigate the transient oscillations by simulating different values of $N$ with temporal noise and comparing it without temporal noise for $N=4.0\times 10^4$. The rescaled $|\alpha|^2$ dynamics shows that the transient high-frequency oscillation has a very minimal amplitude when compared with the oscillations induced by the temporal noise. Increasing the number of atoms effectively decreases the noise level and thereby decreasing the amplitude of the oscillations as shown in Fig.~\ref{fig:N_cav_fft}(a). The power spectra in Fig.~\ref{fig:N_cav_fft}(b) highlight that the inclusion of temporal noise effectively pushes your system to exhibit oscillations close to the expected $\approx \omega_\mathrm{LC}$ even though $\epsilon < \epsilon_\mathrm{LC, MF}$. 

\begin{figure*}[!htb]
	\centering
	\includegraphics[width=1.8\columnwidth]{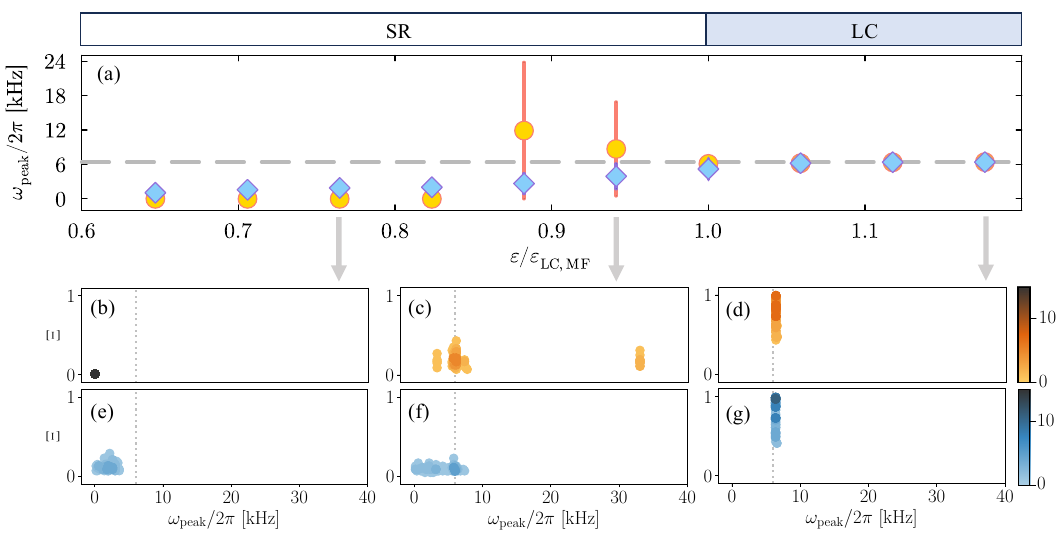}
	\vspace{-0.25cm}
	\caption{(a) The mean frequencies with the standard deviation indicated by the error bars for with (diamond) and without (circle) temporal noise for a one-dimensional atom-cavity system when $\delta_\mathrm{eff} = -2\pi\times 7$ kHz. The shot-to-shot frequency with the relative crystallization fraction for (b-d) with and (e-g) without noise. The horizontal dashed line in (a) and the vertical dotted lines in (b)-(g) correspond to the mean-field limit cycle frequency, $\omega_\mathrm{LC}=2\pi\times 6.0$ kHz, for the closest limit cycle phase.}
	\label{fig:noise1} 
\end{figure*}
\begin{figure*}[!htb]
	\centering
	\includegraphics[width=1.8\columnwidth]{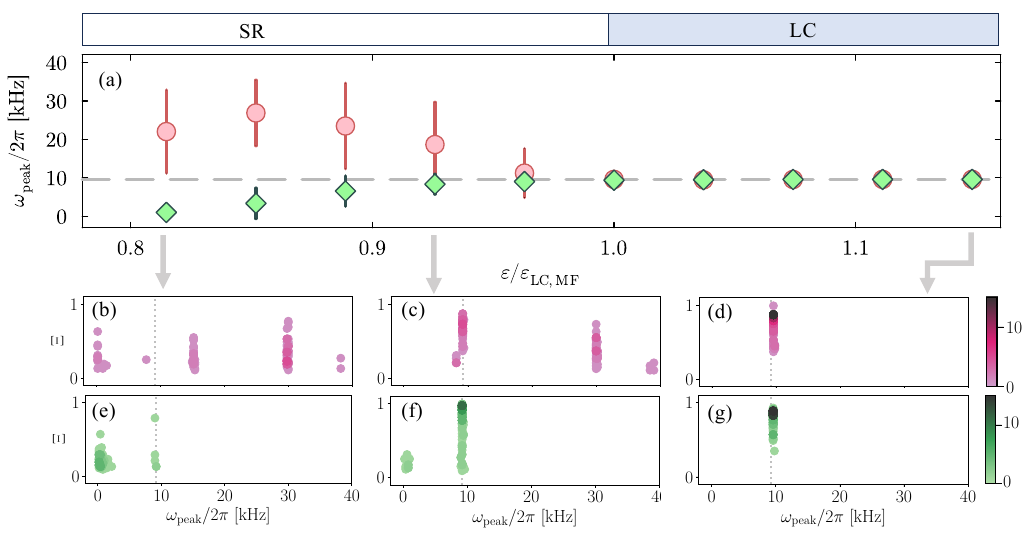}
	\vspace{-0.25cm}
	\caption{(a) Similar to Fig.~5 but for a 2D atom-cavity system with $\delta_\mathrm{eff} = -2\pi\times 4.7$ kHz, and $\omega_\mathrm{LC}=2\pi\times 9.2$ kHz.}
	\label{fig:noise2} 
\end{figure*}

To further explore the noise-induced temporal ordering, we now use 100 trajectories in our TWA simulations for varying pump intensities as depicted in Figs.~\ref{fig:noise1} and~\ref{fig:noise2} for the 1D and 2D systems, respectively. For the pump protocol, we linearly increased the intensity for 5 ms and then kept it constant for 15 ms. We analyze the Fourier transform of the intracavity photon number for $t\in[15, 20]$ ms, $F\{|\alpha|^2(t)\}= |\tilde{\alpha}|^2(\omega)$.
The dominant frequency $\omega_\mathrm{peak}$ is determined as the frequency corresponding to the maximum value of the normalized Fourier spectrum given by the crystalline fraction defined as $\Xi = \max\{|\tilde{\alpha}|^2(\omega)/\sum_\omega |\tilde{\alpha}|^2(\omega)\}$ \cite{Cosme2023}. 

In Figs.~\ref{fig:noise1} and~\ref{fig:noise2}, we present two sets of results depending on which type of quantum noise is included in the TWA. In both cases, quantum noise in the initial state is included. However, in one case, the temporal noise is absent meaning the dynamics is fully deterministic, while in the other, it is present resulting to a stochastic dynamics. Fig.~\ref{fig:noise1} depicts the results for the 1D case while Fig.~\ref{fig:noise2} corresponds to the 2D case.
In Figs.~\ref{fig:noise1}(a) and~\ref{fig:noise2}(a), the average values of $\omega_\mathrm{peak}$ for 100 trajectories are indicated as circles and diamonds for those with and without temporal noise, respectively. The error bars denote the standard deviation for the recorded peak frequencies. In Figs.~\ref{fig:noise1}(b)-(g) and~\ref{fig:noise2}(b)-(g), we present the histograms of the crystalline fraction and peak frequency for exemplary choices of the pump intensity. 

The SR phase manifests as a steady state with zero oscillations in MF as shown in Fig.~\ref{fig:noise1}(b), and low-frequency oscillation due to temporal noise in Fig.~\ref{fig:noise1}(e). 
Trajectories with significantly high frequency in the absence of temporal noise are classified as an SR phase instead of a LC phase as discussed in Sec.~\ref{sec:1D2D}. These low amplitude high-frequency oscillations emerges the closer the pump intensity is to $\varepsilon_\mathrm{LC, MF}$. In some trajectories, these high frequencies have a well-defined oscillation as noted by a high value of $\Xi$ in Fig.~\ref{fig:noise2}(b) and~\ref{fig:noise2}(c). We emphasize that these fast oscillations due to excitations of higher momentum modes are transient as discussed in the previous section, which means that all trajectories in Fig.~\ref{fig:noise2}(b) correspond to an SR phase. Interestingly, Figs.~\ref{fig:noise1}(c),~\ref{fig:noise1}(f),~\ref{fig:noise2}(c), and~\ref{fig:noise2}(f) exhibit trajectories with $\omega_\mathrm{peak}\sim\omega_\mathrm{LC, MF}$ even though $\varepsilon<\varepsilon_\mathrm{LC, MF}$. 
More importantly, comparing Figs.~\ref{fig:noise1}(c) and~\ref{fig:noise1}(f), we find that temporal noise eliminates the presence of trajectories in the SR phase oscillating at high frequencies. This phenomenon is even more pronounced in 2D as seen in Figs.~\ref{fig:noise2}(c) and~\ref{fig:noise2}(f). Hence, the suppression of the high-frequency oscillations in the SR phase by the temporal noise reveals an early signature of the onset of LCs. This noise-induced enhancement of LC is further emphasised in Fig.~\ref{fig:noise2}(f), in which most trajectories possess $\omega_\mathrm{peak}\sim\omega_\mathrm{LC, MF}$. Therefore, our results for the atom-cavity system are consistent with our predictions from the minimal model in Sec.~\ref{sec:modelsims} as summarized in Fig.~\ref{fig:minmodel}.

In general, a CTC is a many-body state, while a LC is an oscillating state in dissipative and nonlinear complex systems, which also include single-particle systems. In addition, not all LCs necessarily correspond to a CTCs due to the requirement of robustness of a CTC, i.e., all trajectories sampling the quantum noise should exhibit a LC response. In the atom-cavity system considered here, the system operating close to the SR-LC transition does not satisfy this requirement as some of the trajectories lead to an SR phase. This can be explained by how the quantum noise in the initial state effectively shifts the phase boundaries because the fluctuating $N_a$ causes slight variations of $N_aU_0$ for constant $U_0$. That is, quantum fluctuations in the initial state effectively changes the $\delta_\mathrm{eff}$, and therefore, the trajectories trajectories would either be an SR or LC near the SR-LC transition.

\subsection{Entrainment}\label{sec:entrainment}

Entrainment, otherwise known as injection locking in optics, is observed by periodically driving a system with a limit cycle state \cite{Strogatz_nonlinear_2019}. This results to the ``locking" of the limit cycle frequency by driving the system to a ratio of $\omega_\mathrm{LC}$ \cite{Jensen1983, Jensen1984}. In the atom-cavity system, this was proposed to create a more robust time crystal or limit cycle phase \cite{Kessler2019,Kessler2020} and was later reported experimentally \cite{kongkhambut2024}. In both of these papers, they focused on subharmonic injection locking, wherein the pump intensity is periodically driven at a frequency $\omega_\mathrm{dr}= 2\omega_\mathrm{LC}$. The effectiveness of entrainment hinges on the periodic driving being as close as possible to an integer multiple of the LC frequency. We now explore the effects of quantum noise on the entrainment in the atom-cavity system.

\begin{figure}[!htpb]
	\centering
	\includegraphics[width=0.97\columnwidth]{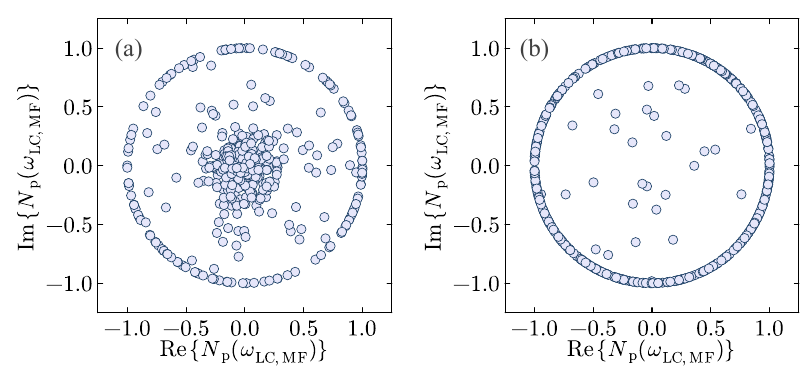}
	\vspace{-0.25cm}
	\caption{Distribution of the time phase in the limit cycle phase for 500 trajectories within TWA for the blue-detuned (a) 1D and (b) 2D atom-cavity system. We use (a) $\varepsilon/\varepsilon_\mathrm{LC, MF} \approx 1.67$, $\omega_\mathrm{LC, MF} = 2\pi\times 8$ kHz, and $\delta_\mathrm{eff} = -2\pi\times7$ kHz and (b) $\varepsilon/\varepsilon_\mathrm{LC, MF} \approx 1.59$, $\omega_\mathrm{LC, MF} =2\pi\times 10.6$ kHz, and $\delta_\mathrm{eff} = -2\pi\times7.4$ kHz.}
	\label{fig:TWA} 
\end{figure} 
First, we explore the shot-to-shot fluctuations deep in the LC regime. To this end, we utilize TWA for 500 trajectories as shown in Fig.~\ref{fig:TWA}. We use the same pump protocol as in the mean-field simulations and kept it constant for 20 ms and 35 ms for 1D and 2D, respectively. We calculate the Fourier transform of the cavity occupation dynamics for each trajectory over the last 5 ms, which means that our Fourier transform can only detect frequency changes larger than 200 Hz. This is important to note when we consider multiple trajectories within TWA as they could in principle have different $\omega_\mathrm{LC}$. From the Fourier spectrum, we record its real and imaginary parts at the LC frequency predicted by MF, $N_P(\omega_\mathrm{LC,MF})$, for fixed  $\delta_\mathrm{eff}$ and $\epsilon$. Furthermore, we rescale the spectrum by the maximum value recorded for all trajectories. Notice that some of the trajectories do not have the same LC frequency as the mean-field prediction, which is evident from the points scattered near the origin of Figs.~\ref{fig:TWA}(a) and~\ref{fig:TWA}(b). Specifically, we note that some of the trajectories have frequency around $\omega_\mathrm{peak} = \omega_{\mathrm{LC,MF}} \pm 2\pi \times 200~\mathrm{Hz}$. This can be explained by the smooth variation of the LC frequency in the mean-field phase diagram. Quantum fluctuations in the initial state lead to shot-to-shot variations of the total particle number $N_a$, which means that TWA can be considered as a sampling of MF results from slightly different choices of $\delta_\mathrm{eff}$ in the mean-field phase diagram. This then leads to different LC frequency for each trajectory, the variation of which depends on how fast the LC frequency changes as a function of $\delta_\mathrm{eff}$. In comparison to Fig.~\ref{fig:TWA}(a), we find that there are less fluctuations of the LC frequency in the 2D system as shown in  Fig.~\ref{fig:TWA}(b), wherein most trajectories are found along the unit circle.

Next, we choose a set of parameters deep in the LC regime with frequency $\omega_\mathrm{LC, MF}$. Then, we periodically modulate the pump intensity once we have a stable LC by driving it at twice the LC frequency $\epsilon=\epsilon_0[1+f_0\cos(2\omega_\mathrm{LC, MF}t)]$. A successful entrainment is marked by $|\alpha|^2$ oscillating at $\omega_\mathrm{LC, MF}$, thereby breaking the discrete time translation symmetry imposed by the periodic drive. This constitutes a phase transition from continuous to discrete time crystals \cite{kongkhambut2024}. 
\begin{figure}[!htb]
	\centering
	\includegraphics[width=0.95\columnwidth]{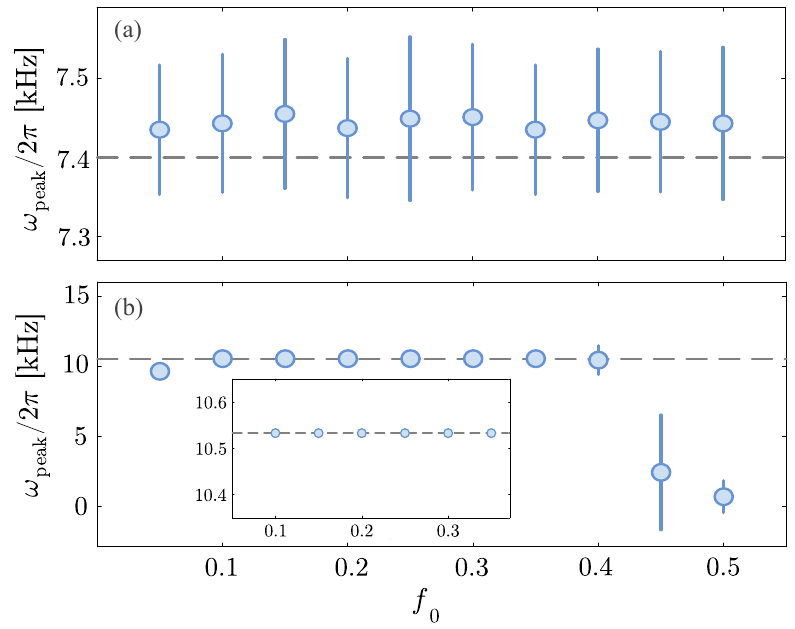}
	\vspace{-0.5cm}
	\caption{Response frequencies as a function of the driving strength $f_0$ for (a) the 1D system with $\omega_\mathrm{LC, MF}=2\pi\times 7.4$ kHz, $\epsilon_0 = 0.28$, and $\delta_\mathrm{eff} = -2\pi\times 7$ kHz, and (b) the 2D system with $\omega_\mathrm{LC, MF} =2\pi\times 10.533$ kHz, $\epsilon_0 = 0.7375$, and $\delta_\mathrm{eff} = -2\pi\times 7.4$ kHz. The driving frequency is $\omega_\mathrm{dr} = 2\omega_\mathrm{LC, MF}$ The error bars denote the standard deviation of the response frequencies obtained for 100 trajectories within TWA.}
	\label{fig:lock} 
\end{figure} 
In Fig.~\ref{fig:lock}, we present the response frequencies for 100 trajectories for different driving strengths $f_0$. The mean value of the response frequency $\omega_\mathrm{peak}$ is indicated as a circle and its standard deviation is represented by the error bar. The dashed lines refer to the LC frequency according to MF, $\omega_\mathrm{LC, MF}$. We observe in Fig.~\ref{fig:lock}(a) that deviations in the response frequencies for all $f_0$ is present. Hence, the entrainment in a 1D system is not as robust as the one observed in Ref.~\cite{kongkhambut2024}.
The weak entrainment can be attributed to the strong shot-to-shot variations of the LC frequency in 1D as demonstrated in Fig.~\ref{fig:TWA}(a). While $\omega_\mathrm{LC,MF}$ is fluctuating for each trajectory, the driving frequency is fixed to just one value for all those trajectories. That is, not all trajectories are resonantly driven, leading to deviations of the response frequency even for smaller values of $f_0$ in 1D. In contrast, the 2D case exhibits a more robust entrainment with zero standard deviation for $f_0\in [0.1, 0.35]$. This demonstrates that, while a 2D system also experiences shot-to-shot fluctuations of the LC frequency (see Fig.~\ref{fig:TWA}(b)), albeit less likely and weaker than the 1D case, it can still result to a successful entrainment as seen in Fig.~\ref{fig:lock}(b). This is further corroborated by a comparison of the 1D case in Fig.~\ref{fig:lock}(a) and the zoomed-in view of the results for 2D in the inset of Fig.~\ref{fig:lock}(b), wherein the standard deviations are considerably higher in 1D than in 2D.

\section{Conclusions}\label{sec:conc}
 {
In summary, we have explored the role of quantum noise in the phase transition involving the emergence of LCs in a class of dissipative systems. 
By considering a generic dissipative model, we show that the transition between a stationary phase or a fixed point and a limit cycle smoothens leading to a noise-induced emergence of LCs. We expect our results to apply to a broad class of systems and here, we demonstrate it for a transversely pumped atom-cavity platform. In particular, we have compared the ensuing photon dynamics according to a MF theory and TWA for the 1D and 2D regimes of the atom-cavity system. The TWA allows us to freely switch on and off the temporal noise associated with the preservation of the bosonic commutation relation in the presence of dissipation. We found that the temporal noise results to an apparent crossover behavior leading to an earlier onset of LCs for smaller pump intensities or, equivalently, light-matter interaction strengths, as compared with the corresponding mean-field prediction. In particular, temporal noise suppresses the high-frequency oscillations in the SR phase, which in turn reveals an earlier signal for the onset of LCs through weak yet detectable oscillations at the expected LC frequency. Furthermore, we have demonstrated that the shot-to-shot fluctuations of the LC frequency are larger in 1D than in the 2D regime of the atom-cavity system. This leads to a less robust entrainment in 1D as some of the trajectories oscillate at frequencies that are not resonant with the applied periodic drive of the pump intensity. 
}

Our work highlights the unique role that quantum noise plays in the emergence and stability of dynamical phases, such as limit cycles and time crystals. In particular, we have shown that the fluctuations corresponding to the photon dissipation, a necessary ingredient for beyond mean-field approaches in open quantum systems to ensure that the bosonic commutation relations for the photons are preserved, may reveal a temporal order, or some hints of it, that is otherwise hidden or completely absent in a standard mean-field theory. Resonance phenomena involving dynamical phases, exemplified in this work by entrainment of LCs, can be adversely affected by quantum fluctuations of the initial state depending on the rigidity of the characteristic frequency of the dynamical phase. It would be interesting to explore in future works these phenomena in the full quantum regime, which may aid in the quest for understanding the quantum origins of various behaviors in nonlinear dynamics, such as the appearance of LCs and bifurcations of stable states \cite{Dutta2024}. The temporal noise in this study can be emulated using a noisy drive appearing as $\approx i\xi(\hat{a}^\dagger - \hat{a})$ in the Hamiltonian of the system. This allows for a careful experimental investigation of noise-induced temporal ordering in dissipative systems that can be described by Eq.~\eqref{eq:minH} in certain regimes.  Alternatively, the atom-cavity setup in Ref.~\cite{Ho2024} can be used for a systematic characterization of the sign-changing LCs that may emerge due to an excitation of a third mode in the effective spin sector. Another interesting future direction would be a careful examination of the apparent transition in the type of LCs between the mesoscopic and mascroscopic regimes either through a detailed semiclassical analysis or a full quantum-mechanical treatment by looking at the Liouvillian and oscillating-mode gaps for varying particle number \cite{Seibold2020,Minganti2020,Buca2022,Bakker2022,Li2024,Jager2024,Cabot2024,Haga2024,Dutta2024}.

\section*{Acknowledgements}
J.G.C. thanks C.J. Cosme for useful discussions. R.J.L.T. and J.G.C. acknowledge support from the DOST-ASTI’s COARE high-performance computing facility. This work was funded by the UP System Balik PhD Program (OVPAA-BPhD-2021-04).

\appendix

\setcounter{equation}{0}
\setcounter{table}{0}

\bibliography{biblio}

\begin{thebibliography}{55}%
\makeatletter
\providecommand \@ifxundefined [1]{%
 \@ifx{#1\undefined}
}%
\providecommand \@ifnum [1]{%
 \ifnum #1\expandafter \@firstoftwo
 \else \expandafter \@secondoftwo
 \fi
}%
\providecommand \@ifx [1]{%
 \ifx #1\expandafter \@firstoftwo
 \else \expandafter \@secondoftwo
 \fi
}%
\providecommand \natexlab [1]{#1}%
\providecommand \enquote  [1]{``#1''}%
\providecommand \bibnamefont  [1]{#1}%
\providecommand \bibfnamefont [1]{#1}%
\providecommand \citenamefont [1]{#1}%
\providecommand \href@noop [0]{\@secondoftwo}%
\providecommand \href [0]{\begingroup \@sanitize@url \@href}%
\providecommand \@href[1]{\@@startlink{#1}\@@href}%
\providecommand \@@href[1]{\endgroup#1\@@endlink}%
\providecommand \@sanitize@url [0]{\catcode `\\12\catcode `\$12\catcode `\&12\catcode `\#12\catcode `\^12\catcode `\_12\catcode `\%12\relax}%
\providecommand \@@startlink[1]{}%
\providecommand \@@endlink[0]{}%
\providecommand \url  [0]{\begingroup\@sanitize@url \@url }%
\providecommand \@url [1]{\endgroup\@href {#1}{\urlprefix }}%
\providecommand \urlprefix  [0]{URL }%
\providecommand \Eprint [0]{\href }%
\providecommand \doibase [0]{https://doi.org/}%
\providecommand \selectlanguage [0]{\@gobble}%
\providecommand \bibinfo  [0]{\@secondoftwo}%
\providecommand \bibfield  [0]{\@secondoftwo}%
\providecommand \translation [1]{[#1]}%
\providecommand \BibitemOpen [0]{}%
\providecommand \bibitemStop [0]{}%
\providecommand \bibitemNoStop [0]{.\EOS\space}%
\providecommand \EOS [0]{\spacefactor3000\relax}%
\providecommand \BibitemShut  [1]{\csname bibitem#1\endcsname}%
\let\auto@bib@innerbib\@empty
\bibitem [{\citenamefont {Matsumoto}\ and\ \citenamefont {Tsuda}(1983)}]{Matsumoto1983}%
  \BibitemOpen
  \bibfield  {author} {\bibinfo {author} {\bibfnamefont {K.}~\bibnamefont {Matsumoto}}\ and\ \bibinfo {author} {\bibfnamefont {I.}~\bibnamefont {Tsuda}},\ }\bibfield  {title} {\bibinfo {title} {Noise-induced order},\ }\href {https://doi.org/10.1007/BF01010923} {\bibfield  {journal} {\bibinfo  {journal} {Journal of Statistical Physics}\ }\textbf {\bibinfo {volume} {31}},\ \bibinfo {pages} {87} (\bibinfo {year} {1983})}\BibitemShut {NoStop}%
\bibitem [{\citenamefont {Zhou}\ and\ \citenamefont {Kurths}(2002)}]{Zhou2002}%
  \BibitemOpen
  \bibfield  {author} {\bibinfo {author} {\bibfnamefont {C.}~\bibnamefont {Zhou}}\ and\ \bibinfo {author} {\bibfnamefont {J.}~\bibnamefont {Kurths}},\ }\bibfield  {title} {\bibinfo {title} {Noise-{Induced} {Phase} {Synchronization} and {Synchronization} {Transitions} in {Chaotic} {Oscillators}},\ }\href {https://doi.org/10.1103/PhysRevLett.88.230602} {\bibfield  {journal} {\bibinfo  {journal} {Physical Review Letters}\ }\textbf {\bibinfo {volume} {88}},\ \bibinfo {pages} {230602} (\bibinfo {year} {2002})}\BibitemShut {NoStop}%
\bibitem [{\citenamefont {Nicolaou}\ \emph {et~al.}(2020)\citenamefont {Nicolaou}, \citenamefont {Sebek}, \citenamefont {Kiss},\ and\ \citenamefont {Motter}}]{Nicolaou2020}%
  \BibitemOpen
  \bibfield  {author} {\bibinfo {author} {\bibfnamefont {Z.~G.}\ \bibnamefont {Nicolaou}}, \bibinfo {author} {\bibfnamefont {M.}~\bibnamefont {Sebek}}, \bibinfo {author} {\bibfnamefont {I.~Z.}\ \bibnamefont {Kiss}},\ and\ \bibinfo {author} {\bibfnamefont {A.~E.}\ \bibnamefont {Motter}},\ }\bibfield  {title} {\bibinfo {title} {Coherent {Dynamics} {Enhanced} by {Uncorrelated} {Noise}},\ }\href {https://doi.org/10.1103/PhysRevLett.125.094101} {\bibfield  {journal} {\bibinfo  {journal} {Physical Review Letters}\ }\textbf {\bibinfo {volume} {125}},\ \bibinfo {pages} {094101} (\bibinfo {year} {2020})}\BibitemShut {NoStop}%
\bibitem [{\citenamefont {Moon}\ \emph {et~al.}(2024)\citenamefont {Moon}, \citenamefont {Schindler}, \citenamefont {Sun}, \citenamefont {Druga}, \citenamefont {Knolle}, \citenamefont {Moessner}, \citenamefont {Zhao}, \citenamefont {Bukov},\ and\ \citenamefont {Ajoy}}]{Moon2024}%
  \BibitemOpen
  \bibfield  {author} {\bibinfo {author} {\bibfnamefont {L.~J.~I.}\ \bibnamefont {Moon}}, \bibinfo {author} {\bibfnamefont {P.~M.}\ \bibnamefont {Schindler}}, \bibinfo {author} {\bibfnamefont {Y.}~\bibnamefont {Sun}}, \bibinfo {author} {\bibfnamefont {E.}~\bibnamefont {Druga}}, \bibinfo {author} {\bibfnamefont {J.}~\bibnamefont {Knolle}}, \bibinfo {author} {\bibfnamefont {R.}~\bibnamefont {Moessner}}, \bibinfo {author} {\bibfnamefont {H.}~\bibnamefont {Zhao}}, \bibinfo {author} {\bibfnamefont {M.}~\bibnamefont {Bukov}},\ and\ \bibinfo {author} {\bibfnamefont {A.}~\bibnamefont {Ajoy}},\ }\href {https://arxiv.org/abs/2404.05620} {\bibinfo {title} {{Experimental observation of a time rondeau crystal: Temporal Disorder in Spatiotemporal Order}}} (\bibinfo {year} {2024}),\ \Eprint {https://arxiv.org/abs/2404.05620} {arXiv:2404.05620 [quant-ph]} \BibitemShut {NoStop}%
\bibitem [{\citenamefont {Zhao}\ \emph {et~al.}(2023)\citenamefont {Zhao}, \citenamefont {Knolle},\ and\ \citenamefont {Moessner}}]{Zhao2023}%
  \BibitemOpen
  \bibfield  {author} {\bibinfo {author} {\bibfnamefont {H.}~\bibnamefont {Zhao}}, \bibinfo {author} {\bibfnamefont {J.}~\bibnamefont {Knolle}},\ and\ \bibinfo {author} {\bibfnamefont {R.}~\bibnamefont {Moessner}},\ }\bibfield  {title} {\bibinfo {title} {Temporal disorder in spatiotemporal order},\ }\href {https://doi.org/10.1103/PhysRevB.108.L100203} {\bibfield  {journal} {\bibinfo  {journal} {Phys. Rev. B}\ }\textbf {\bibinfo {volume} {108}},\ \bibinfo {pages} {L100203} (\bibinfo {year} {2023})}\BibitemShut {NoStop}%
\bibitem [{\citenamefont {Kongkhambut}\ \emph {et~al.}(2022)\citenamefont {Kongkhambut}, \citenamefont {Skulte}, \citenamefont {Mathey}, \citenamefont {Cosme}, \citenamefont {Hemmerich},\ and\ \citenamefont {Keßler}}]{Kongkhambut2022}%
  \BibitemOpen
  \bibfield  {author} {\bibinfo {author} {\bibfnamefont {P.}~\bibnamefont {Kongkhambut}}, \bibinfo {author} {\bibfnamefont {J.}~\bibnamefont {Skulte}}, \bibinfo {author} {\bibfnamefont {L.}~\bibnamefont {Mathey}}, \bibinfo {author} {\bibfnamefont {J.~G.}\ \bibnamefont {Cosme}}, \bibinfo {author} {\bibfnamefont {A.}~\bibnamefont {Hemmerich}},\ and\ \bibinfo {author} {\bibfnamefont {H.}~\bibnamefont {Keßler}},\ }\bibfield  {title} {\bibinfo {title} {Observation of a continuous time crystal},\ }\href {https://doi.org/10.1126/science.abo3382} {\bibfield  {journal} {\bibinfo  {journal} {Science}\ }\textbf {\bibinfo {volume} {377}},\ \bibinfo {pages} {670} (\bibinfo {year} {2022})}\BibitemShut {NoStop}%
\bibitem [{\citenamefont {Heugel}\ \emph {et~al.}(2023)\citenamefont {Heugel}, \citenamefont {Eichler}, \citenamefont {Chitra},\ and\ \citenamefont {Zilberberg}}]{Zilberberg2023}%
  \BibitemOpen
  \bibfield  {author} {\bibinfo {author} {\bibfnamefont {T.~L.}\ \bibnamefont {Heugel}}, \bibinfo {author} {\bibfnamefont {A.}~\bibnamefont {Eichler}}, \bibinfo {author} {\bibfnamefont {R.}~\bibnamefont {Chitra}},\ and\ \bibinfo {author} {\bibfnamefont {O.}~\bibnamefont {Zilberberg}},\ }\bibfield  {title} {\bibinfo {title} {{The role of fluctuations in quantum and classical time crystals}},\ }\href {https://doi.org/10.21468/SciPostPhysCore.6.3.053} {\bibfield  {journal} {\bibinfo  {journal} {SciPost Phys. Core}\ }\textbf {\bibinfo {volume} {6}},\ \bibinfo {pages} {053} (\bibinfo {year} {2023})}\BibitemShut {NoStop}%
\bibitem [{\citenamefont {Piazza}\ and\ \citenamefont {Ritsch}(2015)}]{Piazza2015}%
  \BibitemOpen
  \bibfield  {author} {\bibinfo {author} {\bibfnamefont {F.}~\bibnamefont {Piazza}}\ and\ \bibinfo {author} {\bibfnamefont {H.}~\bibnamefont {Ritsch}},\ }\bibfield  {title} {\bibinfo {title} {{Self-Ordered Limit Cycles, Chaos, and Phase Slippage with a Superfluid inside an Optical Resonator}},\ }\href {https://doi.org/10.1103/PhysRevLett.115.163601} {\bibfield  {journal} {\bibinfo  {journal} {Phys. Rev. Lett.}\ }\textbf {\bibinfo {volume} {115}},\ \bibinfo {pages} {163601} (\bibinfo {year} {2015})}\BibitemShut {NoStop}%
\bibitem [{\citenamefont {{Ke{\ss}ler}}\ \emph {et~al.}(2019)\citenamefont {{Ke{\ss}ler}}, \citenamefont {{Cosme}}, \citenamefont {{Hemmerling}}, \citenamefont {{Mathey}},\ and\ \citenamefont {{Hemmerich}}}]{Kessler2019}%
  \BibitemOpen
  \bibfield  {author} {\bibinfo {author} {\bibfnamefont {H.}~\bibnamefont {{Ke{\ss}ler}}}, \bibinfo {author} {\bibfnamefont {J.~G.}\ \bibnamefont {{Cosme}}}, \bibinfo {author} {\bibfnamefont {M.}~\bibnamefont {{Hemmerling}}}, \bibinfo {author} {\bibfnamefont {L.}~\bibnamefont {{Mathey}}},\ and\ \bibinfo {author} {\bibfnamefont {A.}~\bibnamefont {{Hemmerich}}},\ }\bibfield  {title} {\bibinfo {title} {{Emergent limit cycles and time crystal dynamics in an atom-cavity system}},\ }\href {https://doi.org/10.1103/PhysRevA.99.053605} {\bibfield  {journal} {\bibinfo  {journal} {Phys. Rev. A}\ }\textbf {\bibinfo {volume} {99}},\ \bibinfo {pages} {053605} (\bibinfo {year} {2019})}\BibitemShut {NoStop}%
\bibitem [{\citenamefont {{Ke{\ss}ler}}\ \emph {et~al.}(2020)\citenamefont {{Ke{\ss}ler}}, \citenamefont {{Cosme}}, \citenamefont {{Georges}}, \citenamefont {{Mathey}},\ and\ \citenamefont {{Hemmerich}}}]{Kessler2020}%
  \BibitemOpen
  \bibfield  {author} {\bibinfo {author} {\bibfnamefont {H.}~\bibnamefont {{Ke{\ss}ler}}}, \bibinfo {author} {\bibfnamefont {J.~G.}\ \bibnamefont {{Cosme}}}, \bibinfo {author} {\bibfnamefont {C.}~\bibnamefont {{Georges}}}, \bibinfo {author} {\bibfnamefont {L.}~\bibnamefont {{Mathey}}},\ and\ \bibinfo {author} {\bibfnamefont {A.}~\bibnamefont {{Hemmerich}}},\ }\bibfield  {title} {\bibinfo {title} {{From a continuous to a discrete time crystal in a dissipative atom-cavity system}},\ }\href {https://doi.org/10.1088/1367-2630/ab9fc0} {\bibfield  {journal} {\bibinfo  {journal} {New J. Phys.}\ }\textbf {\bibinfo {volume} {22}},\ \bibinfo {eid} {085002} (\bibinfo {year} {2020})}\BibitemShut {NoStop}%
\bibitem [{\citenamefont {Kongkhambut}\ \emph {et~al.}(2024)\citenamefont {Kongkhambut}, \citenamefont {Cosme}, \citenamefont {Skulte}, \citenamefont {Armijos}, \citenamefont {Mathey}, \citenamefont {Hemmerich},\ and\ \citenamefont {Keßler}}]{kongkhambut2024}%
  \BibitemOpen
  \bibfield  {author} {\bibinfo {author} {\bibfnamefont {P.}~\bibnamefont {Kongkhambut}}, \bibinfo {author} {\bibfnamefont {J.~G.}\ \bibnamefont {Cosme}}, \bibinfo {author} {\bibfnamefont {J.}~\bibnamefont {Skulte}}, \bibinfo {author} {\bibfnamefont {M.~A.~M.}\ \bibnamefont {Armijos}}, \bibinfo {author} {\bibfnamefont {L.}~\bibnamefont {Mathey}}, \bibinfo {author} {\bibfnamefont {A.}~\bibnamefont {Hemmerich}},\ and\ \bibinfo {author} {\bibfnamefont {H.}~\bibnamefont {Keßler}},\ }\bibfield  {title} {\bibinfo {title} {Observation of a phase transition from a continuous to a discrete time crystal},\ }\href {https://doi.org/10.1088/1361-6633/ad6585} {\bibfield  {journal} {\bibinfo  {journal} {Reports on Progress in Physics}\ }\textbf {\bibinfo {volume} {87}},\ \bibinfo {pages} {080502} (\bibinfo {year} {2024})}\BibitemShut {NoStop}%
\bibitem [{\citenamefont {Krishna}\ \emph {et~al.}(2023)\citenamefont {Krishna}, \citenamefont {Solanki}, \citenamefont {Hajdu\ifmmode~\check{s}\else \v{s}\fi{}ek},\ and\ \citenamefont {Vinjanampathy}}]{Krishna2023}%
  \BibitemOpen
  \bibfield  {author} {\bibinfo {author} {\bibfnamefont {M.}~\bibnamefont {Krishna}}, \bibinfo {author} {\bibfnamefont {P.}~\bibnamefont {Solanki}}, \bibinfo {author} {\bibfnamefont {M.}~\bibnamefont {Hajdu\ifmmode~\check{s}\else \v{s}\fi{}ek}},\ and\ \bibinfo {author} {\bibfnamefont {S.}~\bibnamefont {Vinjanampathy}},\ }\bibfield  {title} {\bibinfo {title} {{Measurement-Induced Continuous Time Crystals}},\ }\href {https://doi.org/10.1103/PhysRevLett.130.150401} {\bibfield  {journal} {\bibinfo  {journal} {Phys. Rev. Lett.}\ }\textbf {\bibinfo {volume} {130}},\ \bibinfo {pages} {150401} (\bibinfo {year} {2023})}\BibitemShut {NoStop}%
\bibitem [{\citenamefont {Greilich}\ \emph {et~al.}(2024)\citenamefont {Greilich}, \citenamefont {Kopteva}, \citenamefont {Kamenskii}, \citenamefont {Sokolov}, \citenamefont {Korenev},\ and\ \citenamefont {Bayer}}]{Greilich2024}%
  \BibitemOpen
  \bibfield  {author} {\bibinfo {author} {\bibfnamefont {A.}~\bibnamefont {Greilich}}, \bibinfo {author} {\bibfnamefont {N.~E.}\ \bibnamefont {Kopteva}}, \bibinfo {author} {\bibfnamefont {A.~N.}\ \bibnamefont {Kamenskii}}, \bibinfo {author} {\bibfnamefont {P.~S.}\ \bibnamefont {Sokolov}}, \bibinfo {author} {\bibfnamefont {V.~L.}\ \bibnamefont {Korenev}},\ and\ \bibinfo {author} {\bibfnamefont {M.}~\bibnamefont {Bayer}},\ }\bibfield  {title} {\bibinfo {title} {Robust continuous time crystal in an electron--nuclear spin system},\ }\href {https://doi.org/10.1038/s41567-023-02351-6} {\bibfield  {journal} {\bibinfo  {journal} {Nature Physics}\ }\textbf {\bibinfo {volume} {20}},\ \bibinfo {pages} {631} (\bibinfo {year} {2024})}\BibitemShut {NoStop}%
\bibitem [{\citenamefont {Carraro-Haddad}\ \emph {et~al.}(2024)\citenamefont {Carraro-Haddad}, \citenamefont {Chafatinos}, \citenamefont {Kuznetsov}, \citenamefont {Papuccio-Fern{\'a}ndez}, \citenamefont {Reynoso}, \citenamefont {Bruchhausen}, \citenamefont {Biermann}, \citenamefont {Santos}, \citenamefont {Usaj},\ and\ \citenamefont {Fainstein}}]{Carraro-Haddad2024}%
  \BibitemOpen
  \bibfield  {author} {\bibinfo {author} {\bibfnamefont {I.}~\bibnamefont {Carraro-Haddad}}, \bibinfo {author} {\bibfnamefont {D.~L.}\ \bibnamefont {Chafatinos}}, \bibinfo {author} {\bibfnamefont {A.~S.}\ \bibnamefont {Kuznetsov}}, \bibinfo {author} {\bibfnamefont {I.~A.}\ \bibnamefont {Papuccio-Fern{\'a}ndez}}, \bibinfo {author} {\bibfnamefont {A.~A.}\ \bibnamefont {Reynoso}}, \bibinfo {author} {\bibfnamefont {A.}~\bibnamefont {Bruchhausen}}, \bibinfo {author} {\bibfnamefont {K.}~\bibnamefont {Biermann}}, \bibinfo {author} {\bibfnamefont {P.~V.}\ \bibnamefont {Santos}}, \bibinfo {author} {\bibfnamefont {G.}~\bibnamefont {Usaj}},\ and\ \bibinfo {author} {\bibfnamefont {A.}~\bibnamefont {Fainstein}},\ }\bibfield  {title} {\bibinfo {title} {Solid-state continuous time crystal in a polariton condensate with a built-in mechanical clock},\ }\href {https://doi.org/10.1126/science.adn7087} {\bibfield  {journal} {\bibinfo  {journal} {Science}\ }\textbf {\bibinfo {volume} {384}},\ \bibinfo {pages} {995} (\bibinfo {year} {2024})}\BibitemShut {NoStop}%
\bibitem [{\citenamefont {Ritsch}\ \emph {et~al.}(2013)\citenamefont {Ritsch}, \citenamefont {Domokos}, \citenamefont {Brennecke},\ and\ \citenamefont {Esslinger}}]{Ritsch2013}%
  \BibitemOpen
  \bibfield  {author} {\bibinfo {author} {\bibfnamefont {H.}~\bibnamefont {Ritsch}}, \bibinfo {author} {\bibfnamefont {P.}~\bibnamefont {Domokos}}, \bibinfo {author} {\bibfnamefont {F.}~\bibnamefont {Brennecke}},\ and\ \bibinfo {author} {\bibfnamefont {T.}~\bibnamefont {Esslinger}},\ }\bibfield  {title} {\bibinfo {title} {Cold atoms in cavity-generated dynamical optical potentials},\ }\href {https://doi.org/10.1103/RevModPhys.85.553} {\bibfield  {journal} {\bibinfo  {journal} {Rev. Mod. Phys.}\ }\textbf {\bibinfo {volume} {85}},\ \bibinfo {pages} {553} (\bibinfo {year} {2013})}\BibitemShut {NoStop}%
\bibitem [{\citenamefont {Farokh~Mivehvar}\ and\ \citenamefont {Ritsch}(2021)}]{Mivehvar2021}%
  \BibitemOpen
  \bibfield  {author} {\bibinfo {author} {\bibfnamefont {T.~D.}\ \bibnamefont {Farokh~Mivehvar}, \bibfnamefont {Francesco~Piazza}}\ and\ \bibinfo {author} {\bibfnamefont {H.}~\bibnamefont {Ritsch}},\ }\bibfield  {title} {\bibinfo {title} {{Cavity QED with quantum gases: new paradigms in many-body physics}},\ }\href {https://doi.org/10.1080/00018732.2021.1969727} {\bibfield  {journal} {\bibinfo  {journal} {Advances in Physics}\ }\textbf {\bibinfo {volume} {70}},\ \bibinfo {pages} {1} (\bibinfo {year} {2021})}\BibitemShut {NoStop}%
\bibitem [{\citenamefont {Polkovnikov}\ and\ \citenamefont {Wang}(2004)}]{Polkovnikov2004}%
  \BibitemOpen
  \bibfield  {author} {\bibinfo {author} {\bibfnamefont {A.}~\bibnamefont {Polkovnikov}}\ and\ \bibinfo {author} {\bibfnamefont {D.-W.}\ \bibnamefont {Wang}},\ }\bibfield  {title} {\bibinfo {title} {{Effect of Quantum Fluctuations on the Dipolar Motion of Bose-Einstein Condensates in Optical Lattices}},\ }\href {https://doi.org/10.1103/PhysRevLett.93.070401} {\bibfield  {journal} {\bibinfo  {journal} {Phys. Rev. Lett.}\ }\textbf {\bibinfo {volume} {93}},\ \bibinfo {pages} {070401} (\bibinfo {year} {2004})}\BibitemShut {NoStop}%
\bibitem [{\citenamefont {Dagvadorj}\ \emph {et~al.}(2015)\citenamefont {Dagvadorj}, \citenamefont {Fellows}, \citenamefont {Matyja\ifmmode~\acute{s}\else \'{s}\fi{}kiewicz}, \citenamefont {Marchetti}, \citenamefont {Carusotto},\ and\ \citenamefont {Szyma\ifmmode~\acute{n}\else \'{n}\fi{}ska}}]{Dagvadorj2015}%
  \BibitemOpen
  \bibfield  {author} {\bibinfo {author} {\bibfnamefont {G.}~\bibnamefont {Dagvadorj}}, \bibinfo {author} {\bibfnamefont {J.~M.}\ \bibnamefont {Fellows}}, \bibinfo {author} {\bibfnamefont {S.}~\bibnamefont {Matyja\ifmmode~\acute{s}\else \'{s}\fi{}kiewicz}}, \bibinfo {author} {\bibfnamefont {F.~M.}\ \bibnamefont {Marchetti}}, \bibinfo {author} {\bibfnamefont {I.}~\bibnamefont {Carusotto}},\ and\ \bibinfo {author} {\bibfnamefont {M.~H.}\ \bibnamefont {Szyma\ifmmode~\acute{n}\else \'{n}\fi{}ska}},\ }\bibfield  {title} {\bibinfo {title} {{Nonequilibrium Phase Transition in a Two-Dimensional Driven Open Quantum System}},\ }\href {https://doi.org/10.1103/PhysRevX.5.041028} {\bibfield  {journal} {\bibinfo  {journal} {Phys. Rev. X}\ }\textbf {\bibinfo {volume} {5}},\ \bibinfo {pages} {041028} (\bibinfo {year} {2015})}\BibitemShut {NoStop}%
\bibitem [{\citenamefont {Damanet}\ \emph {et~al.}(2019)\citenamefont {Damanet}, \citenamefont {Daley},\ and\ \citenamefont {Keeling}}]{Damanet2019}%
  \BibitemOpen
  \bibfield  {author} {\bibinfo {author} {\bibfnamefont {F.}~\bibnamefont {Damanet}}, \bibinfo {author} {\bibfnamefont {A.~J.}\ \bibnamefont {Daley}},\ and\ \bibinfo {author} {\bibfnamefont {J.}~\bibnamefont {Keeling}},\ }\bibfield  {title} {\bibinfo {title} {{Atom-only descriptions of the driven-dissipative Dicke model}},\ }\href {https://doi.org/10.1103/PhysRevA.99.033845} {\bibfield  {journal} {\bibinfo  {journal} {Phys. Rev. A}\ }\textbf {\bibinfo {volume} {99}},\ \bibinfo {pages} {033845} (\bibinfo {year} {2019})}\BibitemShut {NoStop}%
\bibitem [{\citenamefont {Pi\~neiro Orioli}\ \emph {et~al.}(2022)\citenamefont {Pi\~neiro Orioli}, \citenamefont {Thompson},\ and\ \citenamefont {Rey}}]{Orioli2022}%
  \BibitemOpen
  \bibfield  {author} {\bibinfo {author} {\bibfnamefont {A.}~\bibnamefont {Pi\~neiro Orioli}}, \bibinfo {author} {\bibfnamefont {J.~K.}\ \bibnamefont {Thompson}},\ and\ \bibinfo {author} {\bibfnamefont {A.~M.}\ \bibnamefont {Rey}},\ }\bibfield  {title} {\bibinfo {title} {{Emergent Dark States from Superradiant Dynamics in Multilevel Atoms in a Cavity}},\ }\href {https://doi.org/10.1103/PhysRevX.12.011054} {\bibfield  {journal} {\bibinfo  {journal} {Phys. Rev. X}\ }\textbf {\bibinfo {volume} {12}},\ \bibinfo {pages} {011054} (\bibinfo {year} {2022})}\BibitemShut {NoStop}%
\bibitem [{\citenamefont {Polkovnikov}(2010)}]{Polkovnikov2010}%
  \BibitemOpen
  \bibfield  {author} {\bibinfo {author} {\bibfnamefont {A.}~\bibnamefont {Polkovnikov}},\ }\bibfield  {title} {\bibinfo {title} {Phase space representation of quantum dynamics},\ }\href {https://doi.org/http://dx.doi.org/10.1016/j.aop.2010.02.006} {\bibfield  {journal} {\bibinfo  {journal} {Ann. Phys.}\ }\textbf {\bibinfo {volume} {325}},\ \bibinfo {pages} {1790} (\bibinfo {year} {2010})}\BibitemShut {NoStop}%
\bibitem [{\citenamefont {{Cosme}}\ \emph {et~al.}(2018)\citenamefont {{Cosme}}, \citenamefont {{Georges}}, \citenamefont {{Hemmerich}},\ and\ \citenamefont {{Mathey}}}]{Cosme2018}%
  \BibitemOpen
  \bibfield  {author} {\bibinfo {author} {\bibfnamefont {J.~G.}\ \bibnamefont {{Cosme}}}, \bibinfo {author} {\bibfnamefont {C.}~\bibnamefont {{Georges}}}, \bibinfo {author} {\bibfnamefont {A.}~\bibnamefont {{Hemmerich}}},\ and\ \bibinfo {author} {\bibfnamefont {L.}~\bibnamefont {{Mathey}}},\ }\bibfield  {title} {\bibinfo {title} {{Dynamical Control of Order in a Cavity-BEC System}},\ }\href {https://doi.org/10.1103/PhysRevLett.121.153001} {\bibfield  {journal} {\bibinfo  {journal} {Phys. Rev. Lett.}\ }\textbf {\bibinfo {volume} {121}},\ \bibinfo {pages} {153001} (\bibinfo {year} {2018})}\BibitemShut {NoStop}%
\bibitem [{\citenamefont {Tuquero}\ \emph {et~al.}(2022)\citenamefont {Tuquero}, \citenamefont {Skulte}, \citenamefont {Mathey},\ and\ \citenamefont {Cosme}}]{Tuquero2022}%
  \BibitemOpen
  \bibfield  {author} {\bibinfo {author} {\bibfnamefont {R.~J.~L.}\ \bibnamefont {Tuquero}}, \bibinfo {author} {\bibfnamefont {J.}~\bibnamefont {Skulte}}, \bibinfo {author} {\bibfnamefont {L.}~\bibnamefont {Mathey}},\ and\ \bibinfo {author} {\bibfnamefont {J.~G.}\ \bibnamefont {Cosme}},\ }\bibfield  {title} {\bibinfo {title} {{Dissipative time crystal in an atom-cavity system: Influence of trap and competing interactions}},\ }\href {https://doi.org/10.1103/PhysRevA.105.043311} {\bibfield  {journal} {\bibinfo  {journal} {Phys. Rev. A}\ }\textbf {\bibinfo {volume} {105}},\ \bibinfo {pages} {043311} (\bibinfo {year} {2022})}\BibitemShut {NoStop}%
\bibitem [{\citenamefont {Cosme}\ \emph {et~al.}(2019)\citenamefont {Cosme}, \citenamefont {Skulte},\ and\ \citenamefont {Mathey}}]{Cosme2019}%
  \BibitemOpen
  \bibfield  {author} {\bibinfo {author} {\bibfnamefont {J.~G.}\ \bibnamefont {Cosme}}, \bibinfo {author} {\bibfnamefont {J.}~\bibnamefont {Skulte}},\ and\ \bibinfo {author} {\bibfnamefont {L.}~\bibnamefont {Mathey}},\ }\bibfield  {title} {\bibinfo {title} {{Time crystals in a shaken atom-cavity system}},\ }\href {https://doi.org/10.1103/PhysRevA.100.053615} {\bibfield  {journal} {\bibinfo  {journal} {Phys. Rev. A}\ }\textbf {\bibinfo {volume} {100}},\ \bibinfo {pages} {053615} (\bibinfo {year} {2019})}\BibitemShut {NoStop}%
\bibitem [{\citenamefont {Ho}\ \emph {et~al.}(2024)\citenamefont {Ho}, \citenamefont {Lua}, \citenamefont {Xiang}, \citenamefont {Rusconi}, \citenamefont {Masson}, \citenamefont {Asenjo-Garcia}, \citenamefont {Yan},\ and\ \citenamefont {Stamper-Kurn}}]{Ho2024}%
  \BibitemOpen
  \bibfield  {author} {\bibinfo {author} {\bibfnamefont {J.}~\bibnamefont {Ho}}, \bibinfo {author} {\bibfnamefont {Y.-H.}\ \bibnamefont {Lua}}, \bibinfo {author} {\bibfnamefont {T.}~\bibnamefont {Xiang}}, \bibinfo {author} {\bibfnamefont {C.~C.}\ \bibnamefont {Rusconi}}, \bibinfo {author} {\bibfnamefont {S.~J.}\ \bibnamefont {Masson}}, \bibinfo {author} {\bibfnamefont {A.}~\bibnamefont {Asenjo-Garcia}}, \bibinfo {author} {\bibfnamefont {Z.}~\bibnamefont {Yan}},\ and\ \bibinfo {author} {\bibfnamefont {D.~M.}\ \bibnamefont {Stamper-Kurn}},\ }\href {https://arxiv.org/abs/2410.12754} {\bibinfo {title} {{Optomechanical self-organization in a mesoscopic atom array}}} (\bibinfo {year} {2024}),\ \Eprint {https://arxiv.org/abs/2410.12754} {arXiv:2410.12754} \BibitemShut {NoStop}%
\bibitem [{\citenamefont {Skulte}\ \emph {et~al.}(2024)\citenamefont {Skulte}, \citenamefont {Kongkhambut}, \citenamefont {Ke\ss{}ler}, \citenamefont {Hemmerich}, \citenamefont {Mathey},\ and\ \citenamefont {Cosme}}]{Skulte2024}%
  \BibitemOpen
  \bibfield  {author} {\bibinfo {author} {\bibfnamefont {J.}~\bibnamefont {Skulte}}, \bibinfo {author} {\bibfnamefont {P.}~\bibnamefont {Kongkhambut}}, \bibinfo {author} {\bibfnamefont {H.}~\bibnamefont {Ke\ss{}ler}}, \bibinfo {author} {\bibfnamefont {A.}~\bibnamefont {Hemmerich}}, \bibinfo {author} {\bibfnamefont {L.}~\bibnamefont {Mathey}},\ and\ \bibinfo {author} {\bibfnamefont {J.~G.}\ \bibnamefont {Cosme}},\ }\bibfield  {title} {\bibinfo {title} {Realizing limit cycles in dissipative bosonic systems},\ }\href {https://doi.org/10.1103/PhysRevA.109.063317} {\bibfield  {journal} {\bibinfo  {journal} {Phys. Rev. A}\ }\textbf {\bibinfo {volume} {109}},\ \bibinfo {pages} {063317} (\bibinfo {year} {2024})}\BibitemShut {NoStop}%
\bibitem [{\citenamefont {Gardiner}\ and\ \citenamefont {Zoller}(2004)}]{Gardiner}%
  \BibitemOpen
  \bibfield  {author} {\bibinfo {author} {\bibfnamefont {C.}~\bibnamefont {Gardiner}}\ and\ \bibinfo {author} {\bibfnamefont {P.}~\bibnamefont {Zoller}},\ }\href@noop {} {\emph {\bibinfo {title} {{Quantum Noise: A Handbook of Markovian and Non-Markovian Quantum Stochastic Methods with Applications to Quantum Optics}}}}\ (\bibinfo  {publisher} {Springer},\ \bibinfo {address} {New York},\ \bibinfo {year} {2004})\BibitemShut {NoStop}%
\bibitem [{\citenamefont {Gardiner}\ and\ \citenamefont {Collett}(1985)}]{Gardiner1986}%
  \BibitemOpen
  \bibfield  {author} {\bibinfo {author} {\bibfnamefont {C.~W.}\ \bibnamefont {Gardiner}}\ and\ \bibinfo {author} {\bibfnamefont {M.~J.}\ \bibnamefont {Collett}},\ }\bibfield  {title} {\bibinfo {title} {{Input and output in damped quantum systems: Quantum stochastic differential equations and the master equation}},\ }\href {https://doi.org/10.1103/PhysRevA.31.3761} {\bibfield  {journal} {\bibinfo  {journal} {Phys. Rev. A}\ }\textbf {\bibinfo {volume} {31}},\ \bibinfo {pages} {3761} (\bibinfo {year} {1985})}\BibitemShut {NoStop}%
\bibitem [{\citenamefont {Clerk}\ \emph {et~al.}(2010)\citenamefont {Clerk}, \citenamefont {Devoret}, \citenamefont {Girvin}, \citenamefont {Marquardt},\ and\ \citenamefont {Schoelkopf}}]{Clerk2010}%
  \BibitemOpen
  \bibfield  {author} {\bibinfo {author} {\bibfnamefont {A.~A.}\ \bibnamefont {Clerk}}, \bibinfo {author} {\bibfnamefont {M.~H.}\ \bibnamefont {Devoret}}, \bibinfo {author} {\bibfnamefont {S.~M.}\ \bibnamefont {Girvin}}, \bibinfo {author} {\bibfnamefont {F.}~\bibnamefont {Marquardt}},\ and\ \bibinfo {author} {\bibfnamefont {R.~J.}\ \bibnamefont {Schoelkopf}},\ }\bibfield  {title} {\bibinfo {title} {Introduction to quantum noise, measurement, and amplification},\ }\href {https://doi.org/10.1103/RevModPhys.82.1155} {\bibfield  {journal} {\bibinfo  {journal} {Rev. Mod. Phys.}\ }\textbf {\bibinfo {volume} {82}},\ \bibinfo {pages} {1155} (\bibinfo {year} {2010})}\BibitemShut {NoStop}%
\bibitem [{\citenamefont {Carusotto}\ and\ \citenamefont {Ciuti}(2013)}]{Carusotto2013}%
  \BibitemOpen
  \bibfield  {author} {\bibinfo {author} {\bibfnamefont {I.}~\bibnamefont {Carusotto}}\ and\ \bibinfo {author} {\bibfnamefont {C.}~\bibnamefont {Ciuti}},\ }\bibfield  {title} {\bibinfo {title} {Quantum fluids of light},\ }\href {https://doi.org/10.1103/RevModPhys.85.299} {\bibfield  {journal} {\bibinfo  {journal} {Rev. Mod. Phys.}\ }\textbf {\bibinfo {volume} {85}},\ \bibinfo {pages} {299} (\bibinfo {year} {2013})}\BibitemShut {NoStop}%
\bibitem [{\citenamefont {Blakie}\ \emph {et~al.}(2008)\citenamefont {Blakie}, \citenamefont {Bradley}, \citenamefont {Davis}, \citenamefont {Ballagh},\ and\ \citenamefont {Gardiner}}]{Blakie2008}%
  \BibitemOpen
  \bibfield  {author} {\bibinfo {author} {\bibfnamefont {P.~B.}\ \bibnamefont {Blakie}}, \bibinfo {author} {\bibfnamefont {A.~S.}\ \bibnamefont {Bradley}}, \bibinfo {author} {\bibfnamefont {M.~J.}\ \bibnamefont {Davis}}, \bibinfo {author} {\bibfnamefont {R.~J.}\ \bibnamefont {Ballagh}},\ and\ \bibinfo {author} {\bibfnamefont {C.~W.}\ \bibnamefont {Gardiner}},\ }\bibfield  {title} {\bibinfo {title} {{Dynamics and statistical mechanics of ultra-cold Bose gases using c-field techniques}},\ }\href {https://doi.org/10.1080/00018730802564254} {\bibfield  {journal} {\bibinfo  {journal} {Adv. Phys.}\ }\textbf {\bibinfo {volume} {57}},\ \bibinfo {pages} {363} (\bibinfo {year} {2008})}\BibitemShut {NoStop}%
\bibitem [{\citenamefont {Kirton}\ \emph {et~al.}(2019)\citenamefont {Kirton}, \citenamefont {Roses}, \citenamefont {Keeling},\ and\ \citenamefont {Dalla~Torre}}]{DickeModel}%
  \BibitemOpen
  \bibfield  {author} {\bibinfo {author} {\bibfnamefont {P.}~\bibnamefont {Kirton}}, \bibinfo {author} {\bibfnamefont {M.~M.}\ \bibnamefont {Roses}}, \bibinfo {author} {\bibfnamefont {J.}~\bibnamefont {Keeling}},\ and\ \bibinfo {author} {\bibfnamefont {E.~G.}\ \bibnamefont {Dalla~Torre}},\ }\bibfield  {title} {\bibinfo {title} {{Introduction to the Dicke model: from equilibrium to nonequilibrium, and vice versa}},\ }\href {https://doi.org/10.1002/qute.201800043} {\bibfield  {journal} {\bibinfo  {journal} {Advanced Quantum Technologies}\ }\textbf {\bibinfo {volume} {2}},\ \bibinfo {pages} {1970013} (\bibinfo {year} {2019})}\BibitemShut {NoStop}%
\bibitem [{\citenamefont {Strogatz}(2000)}]{Strogatz_nonlinear_2019}%
  \BibitemOpen
  \bibfield  {author} {\bibinfo {author} {\bibfnamefont {S.~H.}\ \bibnamefont {Strogatz}},\ }\href@noop {} {\emph {\bibinfo {title} {Nonlinear {Dynamics} and {Chaos}: {With} {Applications} to {Physics}, {Biology}, {Chemistry}, and {Engineering}, in Studies in Nonlinearity}}}\ (\bibinfo  {publisher} {Westview},\ \bibinfo {address} {Cambridge},\ \bibinfo {year} {2000})\BibitemShut {NoStop}%
\bibitem [{\citenamefont {Gao}\ \emph {et~al.}(2002)\citenamefont {Gao}, \citenamefont {Tung},\ and\ \citenamefont {Rao}}]{Gao2002}%
  \BibitemOpen
  \bibfield  {author} {\bibinfo {author} {\bibfnamefont {J.~B.}\ \bibnamefont {Gao}}, \bibinfo {author} {\bibfnamefont {W.-w.}\ \bibnamefont {Tung}},\ and\ \bibinfo {author} {\bibfnamefont {N.}~\bibnamefont {Rao}},\ }\bibfield  {title} {\bibinfo {title} {{Noise-Induced Hopf-Bifurcation-Type Sequence and Transition to Chaos in the Lorenz Equations}},\ }\href {https://doi.org/10.1103/PhysRevLett.89.254101} {\bibfield  {journal} {\bibinfo  {journal} {Phys. Rev. Lett.}\ }\textbf {\bibinfo {volume} {89}},\ \bibinfo {pages} {254101} (\bibinfo {year} {2002})}\BibitemShut {NoStop}%
\bibitem [{\citenamefont {Sch\"utz}\ \emph {et~al.}(2015)\citenamefont {Sch\"utz}, \citenamefont {J\"ager},\ and\ \citenamefont {Morigi}}]{Schutz2015}%
  \BibitemOpen
  \bibfield  {author} {\bibinfo {author} {\bibfnamefont {S.}~\bibnamefont {Sch\"utz}}, \bibinfo {author} {\bibfnamefont {S.~B.}\ \bibnamefont {J\"ager}},\ and\ \bibinfo {author} {\bibfnamefont {G.}~\bibnamefont {Morigi}},\ }\bibfield  {title} {\bibinfo {title} {Thermodynamics and dynamics of atomic self-organization in an optical cavity},\ }\href {https://doi.org/10.1103/PhysRevA.92.063808} {\bibfield  {journal} {\bibinfo  {journal} {Phys. Rev. A}\ }\textbf {\bibinfo {volume} {92}},\ \bibinfo {pages} {063808} (\bibinfo {year} {2015})}\BibitemShut {NoStop}%
\bibitem [{\citenamefont {Stitely}\ \emph {et~al.}(2020)\citenamefont {Stitely}, \citenamefont {Masson}, \citenamefont {Giraldo}, \citenamefont {Krauskopf},\ and\ \citenamefont {Parkins}}]{Stitely2020}%
  \BibitemOpen
  \bibfield  {author} {\bibinfo {author} {\bibfnamefont {K.~C.}\ \bibnamefont {Stitely}}, \bibinfo {author} {\bibfnamefont {S.~J.}\ \bibnamefont {Masson}}, \bibinfo {author} {\bibfnamefont {A.}~\bibnamefont {Giraldo}}, \bibinfo {author} {\bibfnamefont {B.}~\bibnamefont {Krauskopf}},\ and\ \bibinfo {author} {\bibfnamefont {S.}~\bibnamefont {Parkins}},\ }\bibfield  {title} {\bibinfo {title} {{Superradiant switching, quantum hysteresis, and oscillations in a generalized Dicke model}},\ }\href {https://doi.org/10.1103/PhysRevA.102.063702} {\bibfield  {journal} {\bibinfo  {journal} {Phys. Rev. A}\ }\textbf {\bibinfo {volume} {102}},\ \bibinfo {pages} {063702} (\bibinfo {year} {2020})}\BibitemShut {NoStop}%
\bibitem [{\citenamefont {{Nie}}\ and\ \citenamefont {{Zheng}}(2024)}]{Nie2024}%
  \BibitemOpen
  \bibfield  {author} {\bibinfo {author} {\bibfnamefont {X.}~\bibnamefont {{Nie}}}\ and\ \bibinfo {author} {\bibfnamefont {W.}~\bibnamefont {{Zheng}}},\ }\href {https://doi.org/10.48550/arXiv.2411.04687} {\bibinfo {title} {{Fate of the spatial-temporal order under quantum fluctuation}}} (\bibinfo {year} {2024}),\ \Eprint {https://arxiv.org/abs/2411.04687} {arXiv:2411.04687} \BibitemShut {NoStop}%
\bibitem [{\citenamefont {{Zhang}}\ \emph {et~al.}(2022)\citenamefont {{Zhang}}, \citenamefont {{Dreon}}, \citenamefont {{Esslinger}}, \citenamefont {{Jaksch}}, \citenamefont {{Buca}},\ and\ \citenamefont {{Donner}}}]{Zhang2022}%
  \BibitemOpen
  \bibfield  {author} {\bibinfo {author} {\bibfnamefont {Z.}~\bibnamefont {{Zhang}}}, \bibinfo {author} {\bibfnamefont {D.}~\bibnamefont {{Dreon}}}, \bibinfo {author} {\bibfnamefont {T.}~\bibnamefont {{Esslinger}}}, \bibinfo {author} {\bibfnamefont {D.}~\bibnamefont {{Jaksch}}}, \bibinfo {author} {\bibfnamefont {B.}~\bibnamefont {{Buca}}},\ and\ \bibinfo {author} {\bibfnamefont {T.}~\bibnamefont {{Donner}}},\ }\href {https://doi.org/10.48550/arXiv.2205.01461} {\bibinfo {title} {{Tunable Non-equilibrium Phase Transitions between Spatial and Temporal Order through Dissipation}}} (\bibinfo {year} {2022}),\ \Eprint {https://arxiv.org/abs/2205.01461} {arXiv:2205.01461} \BibitemShut {NoStop}%
\bibitem [{\citenamefont {{Ke{\ss}ler}}\ \emph {et~al.}(2021)\citenamefont {{Ke{\ss}ler}}, \citenamefont {{Kongkhambut}}, \citenamefont {{Georges}}, \citenamefont {{Mathey}}, \citenamefont {{Cosme}},\ and\ \citenamefont {{Hemmerich}}}]{Kessler2021}%
  \BibitemOpen
  \bibfield  {author} {\bibinfo {author} {\bibfnamefont {H.}~\bibnamefont {{Ke{\ss}ler}}}, \bibinfo {author} {\bibfnamefont {P.}~\bibnamefont {{Kongkhambut}}}, \bibinfo {author} {\bibfnamefont {C.}~\bibnamefont {{Georges}}}, \bibinfo {author} {\bibfnamefont {L.}~\bibnamefont {{Mathey}}}, \bibinfo {author} {\bibfnamefont {J.~G.}\ \bibnamefont {{Cosme}}},\ and\ \bibinfo {author} {\bibfnamefont {A.}~\bibnamefont {{Hemmerich}}},\ }\bibfield  {title} {\bibinfo {title} {{Observation of a Dissipative Time Crystal}},\ }\href {https://doi.org/10.1103/PhysRevLett.127.043602} {\bibfield  {journal} {\bibinfo  {journal} {Phys. Rev. Lett.}\ }\textbf {\bibinfo {volume} {127}},\ \bibinfo {pages} {043602} (\bibinfo {year} {2021})}\BibitemShut {NoStop}%
\bibitem [{\citenamefont {{Baumann}}\ \emph {et~al.}(2010)\citenamefont {{Baumann}}, \citenamefont {{Guerlin}}, \citenamefont {{Brennecke}},\ and\ \citenamefont {{Esslinger}}}]{Baumann2010}%
  \BibitemOpen
  \bibfield  {author} {\bibinfo {author} {\bibfnamefont {K.}~\bibnamefont {{Baumann}}}, \bibinfo {author} {\bibfnamefont {C.}~\bibnamefont {{Guerlin}}}, \bibinfo {author} {\bibfnamefont {F.}~\bibnamefont {{Brennecke}}},\ and\ \bibinfo {author} {\bibfnamefont {T.}~\bibnamefont {{Esslinger}}},\ }\bibfield  {title} {\bibinfo {title} {{Dicke quantum phase transition with a superfluid gas in an optical cavity}},\ }\href {https://doi.org/10.1038/nature09009} {\bibfield  {journal} {\bibinfo  {journal} {Nature}\ }\textbf {\bibinfo {volume} {464}},\ \bibinfo {pages} {1301} (\bibinfo {year} {2010})}\BibitemShut {NoStop}%
\bibitem [{\citenamefont {{Nagy}}\ \emph {et~al.}(2008)\citenamefont {{Nagy}}, \citenamefont {{Szirmai}},\ and\ \citenamefont {{Domokos}}}]{Nagy2008}%
  \BibitemOpen
  \bibfield  {author} {\bibinfo {author} {\bibfnamefont {D.}~\bibnamefont {{Nagy}}}, \bibinfo {author} {\bibfnamefont {G.}~\bibnamefont {{Szirmai}}},\ and\ \bibinfo {author} {\bibfnamefont {P.}~\bibnamefont {{Domokos}}},\ }\bibfield  {title} {\bibinfo {title} {{Self-organization of a Bose-Einstein condensate in an optical cavity}},\ }\href {https://doi.org/10.1140/epjd/e2008-00074-6} {\bibfield  {journal} {\bibinfo  {journal} {Eur. Phys. J. D}\ }\textbf {\bibinfo {volume} {48}},\ \bibinfo {pages} {127} (\bibinfo {year} {2008})}\BibitemShut {NoStop}%
\bibitem [{\citenamefont {{Klinder}}\ \emph {et~al.}(2015)\citenamefont {{Klinder}}, \citenamefont {{Ke{\ss}ler}}, \citenamefont {{Wolke}}, \citenamefont {{Mathey}},\ and\ \citenamefont {{Hemmerich}}}]{Klinder2015}%
  \BibitemOpen
  \bibfield  {author} {\bibinfo {author} {\bibfnamefont {J.}~\bibnamefont {{Klinder}}}, \bibinfo {author} {\bibfnamefont {H.}~\bibnamefont {{Ke{\ss}ler}}}, \bibinfo {author} {\bibfnamefont {M.}~\bibnamefont {{Wolke}}}, \bibinfo {author} {\bibfnamefont {L.}~\bibnamefont {{Mathey}}},\ and\ \bibinfo {author} {\bibfnamefont {A.}~\bibnamefont {{Hemmerich}}},\ }\bibfield  {title} {\bibinfo {title} {{Dynamical phase transition in the open Dicke model}},\ }\href {https://doi.org/10.1073/pnas.1417132112} {\bibfield  {journal} {\bibinfo  {journal} {Proc. Natl. Acad. Sci. USA}\ }\textbf {\bibinfo {volume} {112}},\ \bibinfo {pages} {3290} (\bibinfo {year} {2015})}\BibitemShut {NoStop}%
\bibitem [{\citenamefont {Gao}\ \emph {et~al.}(2023)\citenamefont {Gao}, \citenamefont {Zhou}, \citenamefont {Guo},\ and\ \citenamefont {Luo}}]{Gao2023}%
  \BibitemOpen
  \bibfield  {author} {\bibinfo {author} {\bibfnamefont {P.}~\bibnamefont {Gao}}, \bibinfo {author} {\bibfnamefont {Z.-W.}\ \bibnamefont {Zhou}}, \bibinfo {author} {\bibfnamefont {G.-C.}\ \bibnamefont {Guo}},\ and\ \bibinfo {author} {\bibfnamefont {X.-W.}\ \bibnamefont {Luo}},\ }\bibfield  {title} {\bibinfo {title} {Self-organized limit cycles in red-detuned atom-cavity systems},\ }\href {https://doi.org/10.1103/PhysRevA.107.023311} {\bibfield  {journal} {\bibinfo  {journal} {Phys. Rev. A}\ }\textbf {\bibinfo {volume} {107}},\ \bibinfo {pages} {023311} (\bibinfo {year} {2023})}\BibitemShut {NoStop}%
\bibitem [{\citenamefont {Cosme}\ \emph {et~al.}(2023)\citenamefont {Cosme}, \citenamefont {Skulte},\ and\ \citenamefont {Mathey}}]{Cosme2023}%
  \BibitemOpen
  \bibfield  {author} {\bibinfo {author} {\bibfnamefont {J.~G.}\ \bibnamefont {Cosme}}, \bibinfo {author} {\bibfnamefont {J.}~\bibnamefont {Skulte}},\ and\ \bibinfo {author} {\bibfnamefont {L.}~\bibnamefont {Mathey}},\ }\bibfield  {title} {\bibinfo {title} {Bridging closed and dissipative discrete time crystals in spin systems with infinite-range interactions},\ }\href {https://doi.org/10.1103/PhysRevB.108.024302} {\bibfield  {journal} {\bibinfo  {journal} {Phys. Rev. B}\ }\textbf {\bibinfo {volume} {108}},\ \bibinfo {pages} {024302} (\bibinfo {year} {2023})}\BibitemShut {NoStop}%
\bibitem [{\citenamefont {Jensen}\ \emph {et~al.}(1983)\citenamefont {Jensen}, \citenamefont {Bak},\ and\ \citenamefont {Bohr}}]{Jensen1983}%
  \BibitemOpen
  \bibfield  {author} {\bibinfo {author} {\bibfnamefont {M.~H.}\ \bibnamefont {Jensen}}, \bibinfo {author} {\bibfnamefont {P.}~\bibnamefont {Bak}},\ and\ \bibinfo {author} {\bibfnamefont {T.}~\bibnamefont {Bohr}},\ }\bibfield  {title} {\bibinfo {title} {{Complete Devil's Staircase, Fractal Dimension, and Universality of Mode- Locking Structure in the Circle Map}},\ }\href {https://doi.org/10.1103/PhysRevLett.50.1637} {\bibfield  {journal} {\bibinfo  {journal} {Phys. Rev. Lett.}\ }\textbf {\bibinfo {volume} {50}},\ \bibinfo {pages} {1637} (\bibinfo {year} {1983})}\BibitemShut {NoStop}%
\bibitem [{\citenamefont {Jensen}\ \emph {et~al.}(1984)\citenamefont {Jensen}, \citenamefont {Bak},\ and\ \citenamefont {Bohr}}]{Jensen1984}%
  \BibitemOpen
  \bibfield  {author} {\bibinfo {author} {\bibfnamefont {M.~H.}\ \bibnamefont {Jensen}}, \bibinfo {author} {\bibfnamefont {P.}~\bibnamefont {Bak}},\ and\ \bibinfo {author} {\bibfnamefont {T.}~\bibnamefont {Bohr}},\ }\bibfield  {title} {\bibinfo {title} {{Transition to chaos by interaction of resonances in dissipative systems. I. Circle maps}},\ }\href {https://doi.org/10.1103/PhysRevA.30.1960} {\bibfield  {journal} {\bibinfo  {journal} {Phys. Rev. A}\ }\textbf {\bibinfo {volume} {30}},\ \bibinfo {pages} {1960} (\bibinfo {year} {1984})}\BibitemShut {NoStop}%
\bibitem [{\citenamefont {Dutta}\ \emph {et~al.}(2024)\citenamefont {Dutta}, \citenamefont {Zhang},\ and\ \citenamefont {Haque}}]{Dutta2024}%
  \BibitemOpen
  \bibfield  {author} {\bibinfo {author} {\bibfnamefont {S.}~\bibnamefont {Dutta}}, \bibinfo {author} {\bibfnamefont {S.}~\bibnamefont {Zhang}},\ and\ \bibinfo {author} {\bibfnamefont {M.}~\bibnamefont {Haque}},\ }\href {https://arxiv.org/abs/2405.08866} {\bibinfo {title} {On the quantum origin of limit cycles, fixed points, and critical slowing down}} (\bibinfo {year} {2024}),\ \Eprint {https://arxiv.org/abs/2405.08866} {arXiv:2405.08866} \BibitemShut {NoStop}%
\bibitem [{\citenamefont {Seibold}\ \emph {et~al.}(2020)\citenamefont {Seibold}, \citenamefont {Rota},\ and\ \citenamefont {Savona}}]{Seibold2020}%
  \BibitemOpen
  \bibfield  {author} {\bibinfo {author} {\bibfnamefont {K.}~\bibnamefont {Seibold}}, \bibinfo {author} {\bibfnamefont {R.}~\bibnamefont {Rota}},\ and\ \bibinfo {author} {\bibfnamefont {V.}~\bibnamefont {Savona}},\ }\bibfield  {title} {\bibinfo {title} {Dissipative time crystal in an asymmetric nonlinear photonic dimer},\ }\href {https://doi.org/10.1103/PhysRevA.101.033839} {\bibfield  {journal} {\bibinfo  {journal} {Phys. Rev. A}\ }\textbf {\bibinfo {volume} {101}},\ \bibinfo {pages} {033839} (\bibinfo {year} {2020})}\BibitemShut {NoStop}%
\bibitem [{\citenamefont {{Minganti}}\ \emph {et~al.}(2020)\citenamefont {{Minganti}}, \citenamefont {{Arkhipov}}, \citenamefont {{Miranowicz}},\ and\ \citenamefont {{Nori}}}]{Minganti2020}%
  \BibitemOpen
  \bibfield  {author} {\bibinfo {author} {\bibfnamefont {F.}~\bibnamefont {{Minganti}}}, \bibinfo {author} {\bibfnamefont {I.~I.}\ \bibnamefont {{Arkhipov}}}, \bibinfo {author} {\bibfnamefont {A.}~\bibnamefont {{Miranowicz}}},\ and\ \bibinfo {author} {\bibfnamefont {F.}~\bibnamefont {{Nori}}},\ }\href {https://doi.org/10.48550/arXiv.2008.08075} {\bibinfo {title} {{Correspondence between dissipative phase transitions of light and time crystals}}} (\bibinfo {year} {2020}),\ \Eprint {https://arxiv.org/abs/2008.08075} {arXiv:2008.08075} \BibitemShut {NoStop}%
\bibitem [{\citenamefont {Buča}\ \emph {et~al.}(2022)\citenamefont {Buča}, \citenamefont {Booker},\ and\ \citenamefont {Jaksch}}]{Buca2022}%
  \BibitemOpen
  \bibfield  {author} {\bibinfo {author} {\bibfnamefont {B.}~\bibnamefont {Buča}}, \bibinfo {author} {\bibfnamefont {C.}~\bibnamefont {Booker}},\ and\ \bibinfo {author} {\bibfnamefont {D.}~\bibnamefont {Jaksch}},\ }\bibfield  {title} {\bibinfo {title} {{Algebraic theory of quantum synchronization and limit cycles under dissipation}},\ }\href {https://doi.org/10.21468/SciPostPhys.12.3.097} {\bibfield  {journal} {\bibinfo  {journal} {SciPost Phys.}\ }\textbf {\bibinfo {volume} {12}},\ \bibinfo {pages} {097} (\bibinfo {year} {2022})}\BibitemShut {NoStop}%
\bibitem [{\citenamefont {Bakker}\ \emph {et~al.}(2022)\citenamefont {Bakker}, \citenamefont {Bahovadinov}, \citenamefont {Kurlov}, \citenamefont {Gritsev}, \citenamefont {Fedorov},\ and\ \citenamefont {Krimer}}]{Bakker2022}%
  \BibitemOpen
  \bibfield  {author} {\bibinfo {author} {\bibfnamefont {L.~R.}\ \bibnamefont {Bakker}}, \bibinfo {author} {\bibfnamefont {M.~S.}\ \bibnamefont {Bahovadinov}}, \bibinfo {author} {\bibfnamefont {D.~V.}\ \bibnamefont {Kurlov}}, \bibinfo {author} {\bibfnamefont {V.}~\bibnamefont {Gritsev}}, \bibinfo {author} {\bibfnamefont {A.~K.}\ \bibnamefont {Fedorov}},\ and\ \bibinfo {author} {\bibfnamefont {D.~O.}\ \bibnamefont {Krimer}},\ }\bibfield  {title} {\bibinfo {title} {{Driven-Dissipative Time Crystalline Phases in a Two-Mode Bosonic System with Kerr Nonlinearity}},\ }\href {https://doi.org/10.1103/PhysRevLett.129.250401} {\bibfield  {journal} {\bibinfo  {journal} {Phys. Rev. Lett.}\ }\textbf {\bibinfo {volume} {129}},\ \bibinfo {pages} {250401} (\bibinfo {year} {2022})}\BibitemShut {NoStop}%
\bibitem [{\citenamefont {Li}\ \emph {et~al.}(2024)\citenamefont {Li}, \citenamefont {Wang}, \citenamefont {Tang},\ and\ \citenamefont {Liu}}]{Li2024}%
  \BibitemOpen
  \bibfield  {author} {\bibinfo {author} {\bibfnamefont {Y.}~\bibnamefont {Li}}, \bibinfo {author} {\bibfnamefont {C.}~\bibnamefont {Wang}}, \bibinfo {author} {\bibfnamefont {Y.}~\bibnamefont {Tang}},\ and\ \bibinfo {author} {\bibfnamefont {Y.-C.}\ \bibnamefont {Liu}},\ }\bibfield  {title} {\bibinfo {title} {{Time Crystal in a Single-Mode Nonlinear Cavity}},\ }\href {https://doi.org/10.1103/PhysRevLett.132.183803} {\bibfield  {journal} {\bibinfo  {journal} {Phys. Rev. Lett.}\ }\textbf {\bibinfo {volume} {132}},\ \bibinfo {pages} {183803} (\bibinfo {year} {2024})}\BibitemShut {NoStop}%
\bibitem [{\citenamefont {J\"ager}\ \emph {et~al.}(2024)\citenamefont {J\"ager}, \citenamefont {Giesen}, \citenamefont {Schneider},\ and\ \citenamefont {Eggert}}]{Jager2024}%
  \BibitemOpen
  \bibfield  {author} {\bibinfo {author} {\bibfnamefont {S.~B.}\ \bibnamefont {J\"ager}}, \bibinfo {author} {\bibfnamefont {J.~M.}\ \bibnamefont {Giesen}}, \bibinfo {author} {\bibfnamefont {I.}~\bibnamefont {Schneider}},\ and\ \bibinfo {author} {\bibfnamefont {S.}~\bibnamefont {Eggert}},\ }\bibfield  {title} {\bibinfo {title} {{Dissipative Dicke time crystals: An atom's point of view}},\ }\href {https://doi.org/10.1103/PhysRevA.110.L010202} {\bibfield  {journal} {\bibinfo  {journal} {Phys. Rev. A}\ }\textbf {\bibinfo {volume} {110}},\ \bibinfo {pages} {L010202} (\bibinfo {year} {2024})}\BibitemShut {NoStop}%
\bibitem [{\citenamefont {Cabot}\ \emph {et~al.}(2024)\citenamefont {Cabot}, \citenamefont {Giorgi},\ and\ \citenamefont {Zambrini}}]{Cabot2024}%
  \BibitemOpen
  \bibfield  {author} {\bibinfo {author} {\bibfnamefont {A.}~\bibnamefont {Cabot}}, \bibinfo {author} {\bibfnamefont {G.~L.}\ \bibnamefont {Giorgi}},\ and\ \bibinfo {author} {\bibfnamefont {R.}~\bibnamefont {Zambrini}},\ }\bibfield  {title} {\bibinfo {title} {{Nonequilibrium Transition between Dissipative Time Crystals}},\ }\href {https://doi.org/10.1103/PRXQuantum.5.030325} {\bibfield  {journal} {\bibinfo  {journal} {PRX Quantum}\ }\textbf {\bibinfo {volume} {5}},\ \bibinfo {pages} {030325} (\bibinfo {year} {2024})}\BibitemShut {NoStop}%
\bibitem [{\citenamefont {Haga}(2024)}]{Haga2024}%
  \BibitemOpen
  \bibfield  {author} {\bibinfo {author} {\bibfnamefont {T.}~\bibnamefont {Haga}},\ }\bibfield  {title} {\bibinfo {title} {{Oscillating-mode gap: An indicator of phase transitions in open quantum many-body systems}},\ }\href {https://doi.org/10.1103/PhysRevB.110.104303} {\bibfield  {journal} {\bibinfo  {journal} {Phys. Rev. B}\ }\textbf {\bibinfo {volume} {110}},\ \bibinfo {pages} {104303} (\bibinfo {year} {2024})}\BibitemShut {NoStop}%
\end{thebibliography}%

\end{document}